\documentclass[aps,prd,twocolumn,superscriptaddress]{revtex4}
\usepackage{graphicx}
\usepackage{amsmath,amssymb,latexsym}
\usepackage{bm}
\usepackage{aas_macros} 

\begin{document}

\title{Gamma-ray emission of hot astrophysical plasmas}

\author{Ervin~Kafexhiu}
\affiliation{Max-Planck-Institut f\"ur Kernphysik, Saupfercheckweg 1, D-69117 Heidelberg, Germany}
\email[E-mail us at: ]{ervin.kafexhiu@mpi-hd.mpg.de}

\author{Felix~Aharonian}
\affiliation{Max-Planck-Institut f\"ur Kernphysik, Saupfercheckweg 1, D-69117 Heidelberg, Germany}
\affiliation{Dublin Institute for Advanced Studies, 31 Fitzwilliam Place, Dublin 2, Ireland}
\affiliation{National Research Nuclear University MEPhI, Kashirskoje Shosse, 31, 115409 Moscow, Russia}

\author{Maxim Barkov}
\affiliation{Department of Physics and Astronomy, Purdue University, West Lafayette, IN 47907-2036, USA}
\affiliation{Astrophysical Big Bang Laboratory, RIKEN, 351-0198 Saitama, Japan}
\affiliation{Space Research Institute of the Russian Academy of Sciences (IKI), 84/32 Profsoyuznaya Str, Moscow, Russia, 117997}

\begin{abstract}
Very hot plasmas with ion temperature exceeding $10^{10}$~K can be formed in certain astrophysical environments. The distinct radiation signature of such plasmas is the $\gamma$-ray emission dominated by the prompt de-excitation nuclear lines and $\pi^0$-decay $\gamma$-rays. Using a large nuclear reaction network, we compute the time evolution of the chemical composition of such hot plasmas and their $\gamma$-ray line emissivity.  At higher energies, we provide simple but accurate analytical presentations for the $\pi^0$-meson production rate and the corresponding  $\pi^0\to2\gamma$ emissivity derived for the Maxwellian distribution of protons. We discuss the impact of the possible deviation of the high energy tail of the particle distribution function from the ``nominal'' Maxwellian distribution on the plasma $\gamma$-ray emissivity. 
\end{abstract}

\maketitle

\section{Introduction \label{intro}}

Very hot astrophysical plasmas with ion temperatures $T_{\rm i}>10^{10}$~K ($k T_{\rm i} \gtrsim1$~MeV) can be formed in optically thin accretion flows around compact relativistic objects (see e.g. Refs.~\cite{Shapiro1976, Narayan1994, Blandford1999}) as well as in strong  (sub-relativistic) shock waves linked to extreme astrophysical phenomena such as supernova explosions (see e.g. Ref.~\cite{Colgate1975}). Typically, in such environments, the ion temperature significantly exceeds the electron temperature, $T_{\rm i} \gg  T_{\rm e}$. Due to intense radiative cooling, the electron temperatures are typically well below $10^{10}$~K; the corresponding optically thin or Comptonised free-free emission appears in the hard X-ray band. Ions, on the other hand, can emit prompt $\gamma$-ray lines in the 0.1--10~MeV energy band via the nuclear excitation channel (see e.g. Ref.~\cite{Ramaty1979, Belhout2007, Murphy2009}).

In the interaction of sub-relativistic nuclei with the cold gas, only a small ($\leq 10^{-6}$) fraction of the nucleus kinetic energy is released in $\gamma$-ray lines, while the rest goes to the heating and ionization of the ambient medium. In hot two temperature plasmas with the electron temperature $T_{\rm e} \geq 10^{8}$~K, plasmas, because of reduction of the Coulomb exchange rate between ions and electrons, the ion temperature can grow, and the efficiency of $\gamma$-ray line emission can be increased. On the other hand, at ion temperatures $T >10^{10}$~K, the inelastic ion collisions tend to destroy nuclei. On average, a nucleus can get excited no more than once before its spallation. As a result, the prompt $\gamma$-ray line emission efficiency is considerably reduced to less than 0.1~\% of the total plasma emissivity ~\cite{Aharonian1984}.

If the characteristic timescale of the system (e.g. the accretion time in the inner disk) is larger than the relevant spallation times, then all nuclei are destroyed and a neutron--proton plasma is formed with some  content of deuterium \citep[see e.g.][]{Aharonian1984, Agaronyan1987}. Thus, at that stage of evolution, the emission of nuclear lines disappear. Yet, the neutron--proton plasma emits a continuum $\gamma$-radiation due to the radiative capture of neutrons ($n+p\to D+\gamma$) and the proton-neutron bremsstrahlung \cite{Aharonian1984}.

Because of spallation of nuclei on timescales shorter than the characteristic ``lifetime'' of the hot plasma, the time evolution of chemical composition is essential for accurate calculations of the emissivity of radiation both in prompt $\gamma$-ray lines and in the $p-n$ continuum. Thus, the time evolution of chemical composition of plasma is essential for accurate calculations of the emissivity of radiation both in prompt $\gamma$-ray lines and in the $p-n$ continuum. In the context of high energy radiation of accreting black holes, this issue has been discussed as early as in the 1980s \cite{Aharonian1984}, however so far no detailed calculations have been performed. This could be explained by the little practical interest of the topic because of the faint gamma-ray fluxes which even in the case of realization of optimal conditions of their production, have been out of rich of detectability by the available telescopes. For detection of prompt $\gamma$-ray lines from the most promising black-hole candidates in our Galaxy, such as Cyg X-1, the flux sensitivity of gamma-ray detectors operating in the MeV band should achieve a level as low as $10^{−12}~{\rm  erg/cm^2s}$. The design and construction of telescopes with such a sensitivity remains a challenge for MeV gamma-ray astronomy, but with arrival of the new generation low-energy gamma-ray missions such as eASTROGAM \cite{astrogam} and AMEGO \cite{amego}, a real breakthrough is expected in the field. This was one of the motivations of the study presented in this paper.

At ion temperatures significantly exceeding $10^{10}$~K, the inelastic proton--proton and proton--neutron interactions open a new effective channel of $\gamma$-ray emission through the production and subsequent decay of secondary $\pi^0$-mesons (see e.g. Ref.~\cite{dahlbacka1974, kolykhalov&sunyaev1979, giovannelli1982, AharonianAtoyan1983, dermer1986}). The energy thresholds of excitation of nuclei are around several MeV/nucleon, therefore at such high temperatures the nuclear $\gamma$-ray lines are produced mainly by particles that populate the central part of the Maxwellian distribution of nuclei. The threshold of $\pi$-meson production is much higher, $E_{\rm k}^{\rm th} =2\,m_\pi\,c^2 (1+m_\pi/4 m_p) \approx 280$~MeV. This implies that $\pi^0$-mesons are produced when at least one of the nucleons populates the high energy tail ($E \gg k T_{\rm i}$) of the Maxwellian distribution.

In reality, the short timescales characterizing the lifetime of hot plasma (for example, the accretion time of the flow in the proximity to the event horizon of the black hole) can prevent the plasma from developing a Maxwellian distribution tail. The central region of the Maxwellian distribution around the average particle energy is established in a short timescale, after just a few binary elastic scatterings. However, the formation of the high energy Maxwellian tail requires much longer time, thus in certain accretion regimes the $\gamma$-ray production would be suppressed \cite{AharonianAtoyan1983}). On the other hand, the plasma instabilities can initiate acceleration of particles and form a suprathermal component well above the ``nominal'' Maxwellian distribution. Correspondingly, the $\gamma$-ray emissivity would be significantly enhanced. Both scenarios should have a dramatic impact on the overall $\pi^0$-meson production rate, and therefore on the $\gamma$-ray luminosity of the source. The $\gamma$-ray spectra of plasma with an underdeveloped Maxwellian tail, or with an excess of high energy particles contributed by different acceleration processes, are expected to be entirely different. In the first case, the radiation will be concentrated around 100 MeV and below, while the presence of the suprathermal component of protons would result in a hard $\gamma$-ray spectra extending well beyond 100~MeV.   
The new generation telescopes, like eASTROGAM and AMEGO, designed for effective $\gamma$-ray detection both in the MeV and GeV bands, can serve as effective tools for the identification of these distinct spectral features.

Below, we present the results of calculations of the $\gamma$-radiation of very hot ion plasmas in the MeV to GeV energy band. In these calculations, we include the time evolution of the chemical composition of the plasma and compute the emissivity of the prompt nuclear $\gamma$-ray lines and the $\gamma$-ray continuum. The calculations depend on the ion temperature of plasma, its chemical composition, and density. In low-density plasma, the processes are dominated by binary interactions. We use the methods developed for calculations of the nuclear cross-section \cite{kafexhiu2012} as well as the parametrisations of $\pi^0$-meson production cross sections with particular attention to the energy region close to the kinematic threshold \cite{kafexhiu2014}.

\begin{figure*}
\includegraphics[scale=0.4]{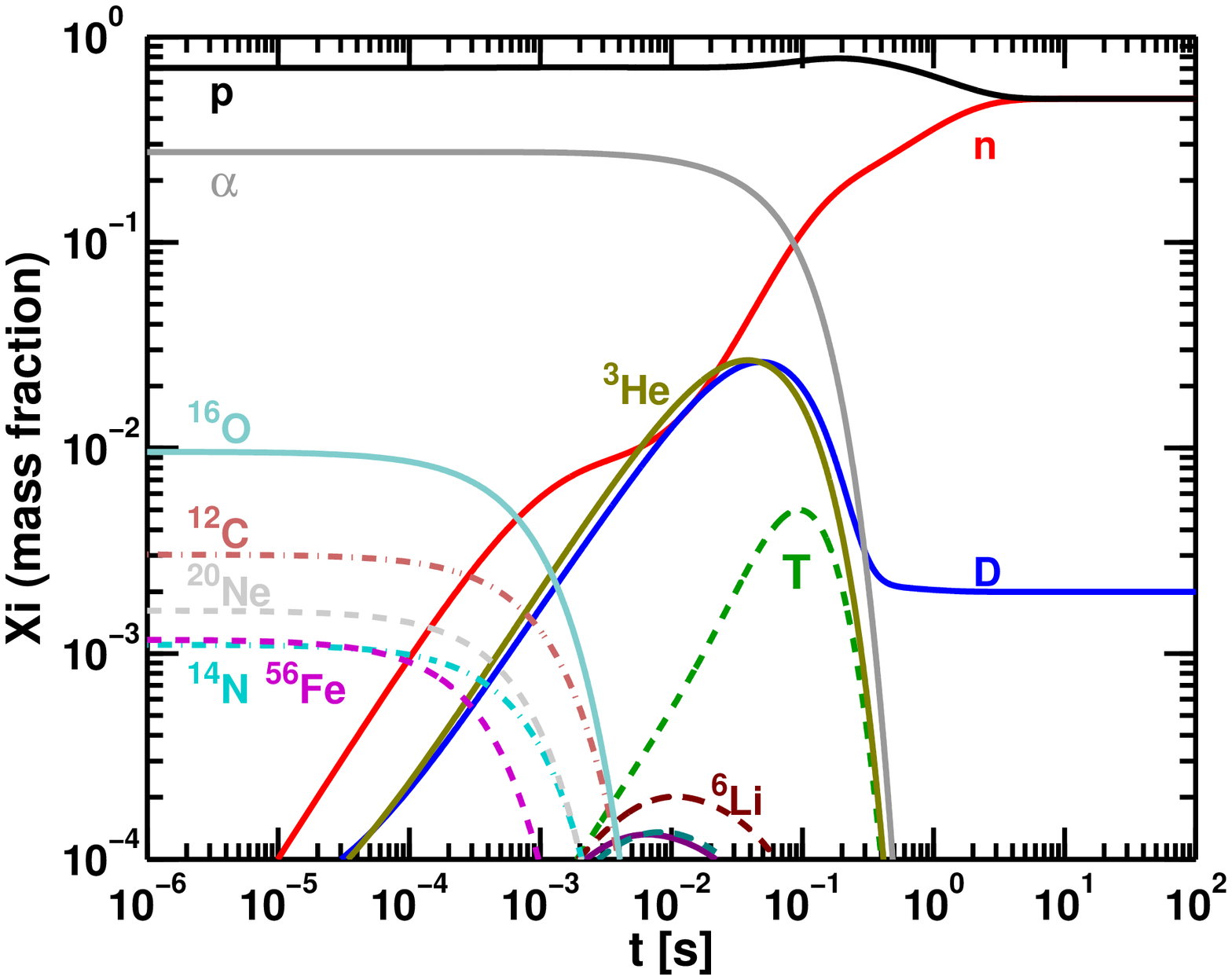}
\includegraphics[scale=0.4]{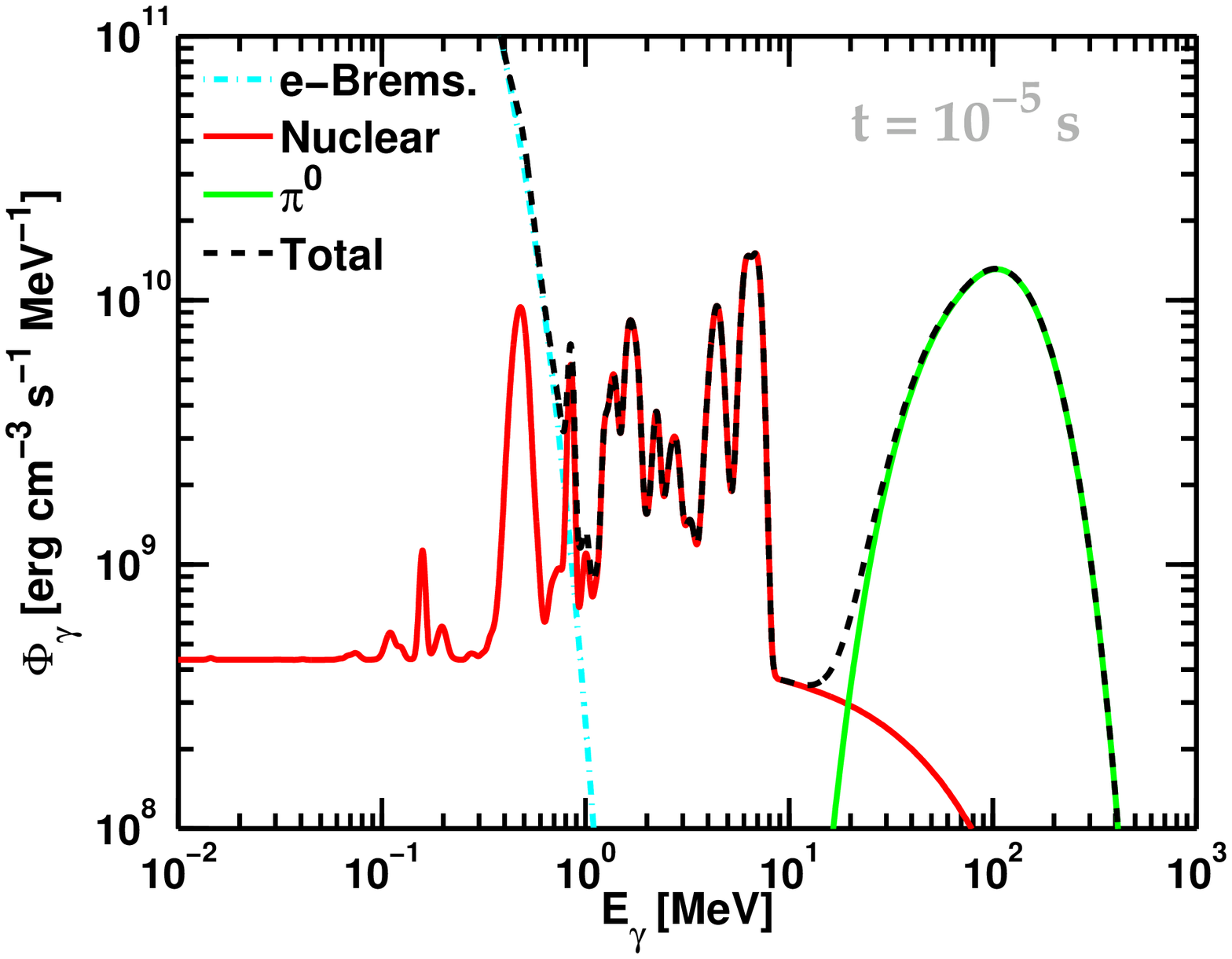}\\
\includegraphics[scale=0.4]{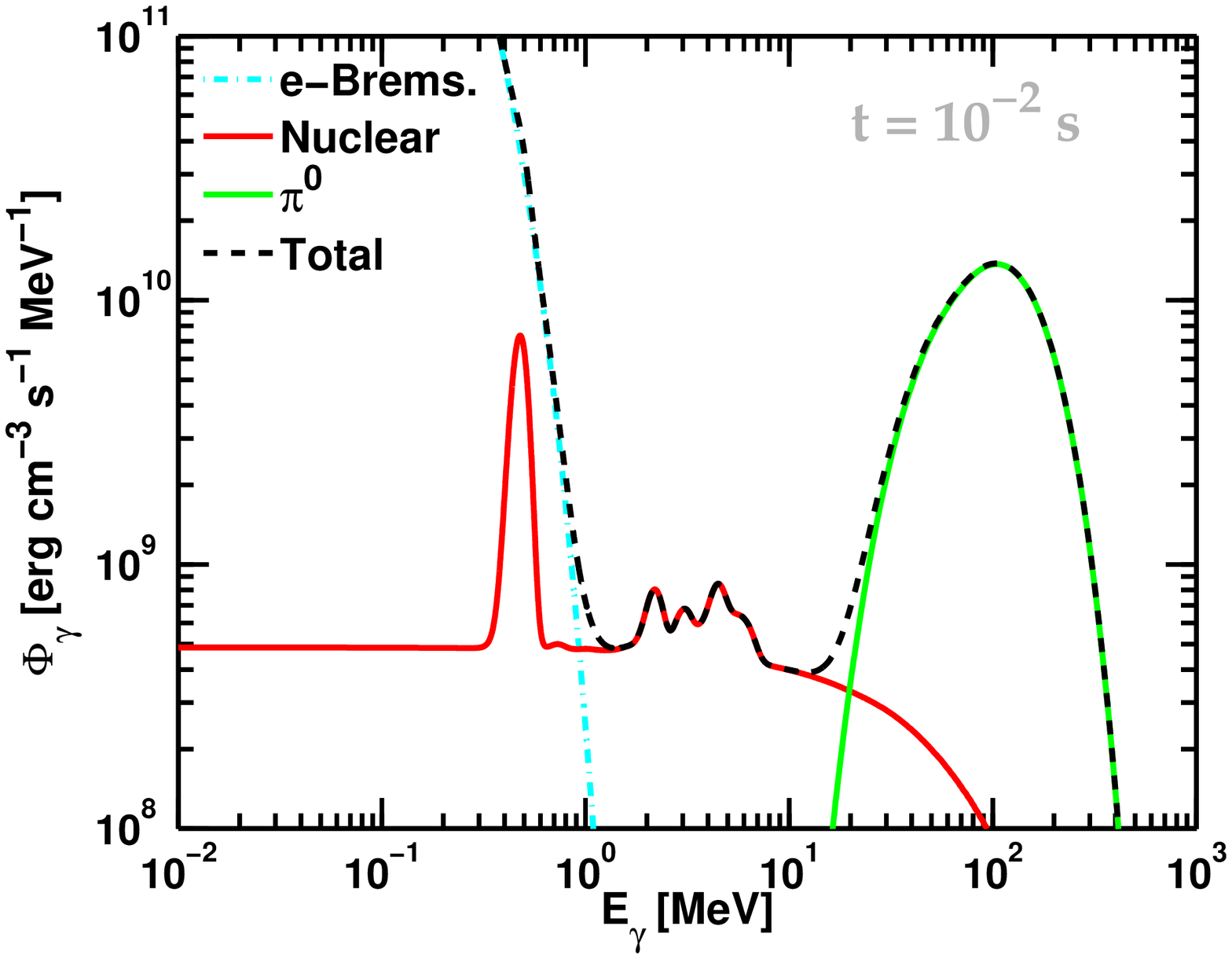}
\includegraphics[scale=0.4]{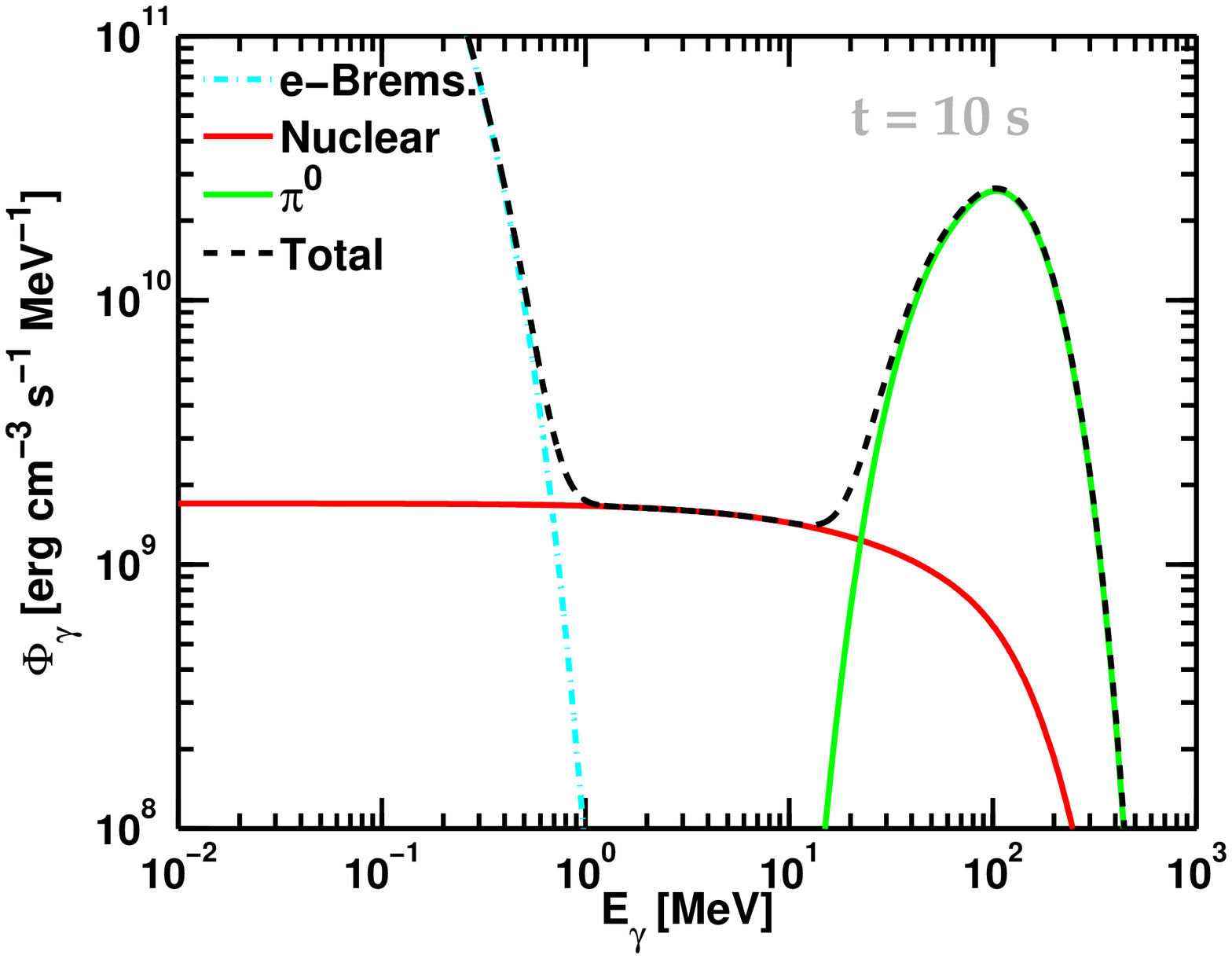}
\caption{Time evolution of the plasma chemical composition and its $\gamma$-ray emissivity. The plasma initial composition is a solar composition and its temperature and nucleon density are fixed to $k T_{\rm i}=50$~MeV and $\rho_{17}=1$. The top left panel shows the temporal chemical evolution of some important elements of the plasma. The other three panels show the plasma $\gamma$-ray emissivity for three specific instants of the plasma evolution: $t=10^{-5}$~s (top right), $t=10^{-2}$~s (bootom left) and $t=10$~s (bottom right). The cyan dash-dot-line shows the thermal electron bremsstrahlung at $k T_{\rm e}=100$~keV (typical for two temperature accretion models). The red line is the $\gamma$-ray emissivity produced by nuclear reactions including nuclear $\gamma$-ray lines and continuum. The green line shows the emissivity due to $\pi^0$-meson production including the contribution from nuclei. The black dash-line is the sum of all emissivities. \label{fig:nucgamspec}}
\end{figure*}

\begin{figure*}
\includegraphics[scale=0.4]{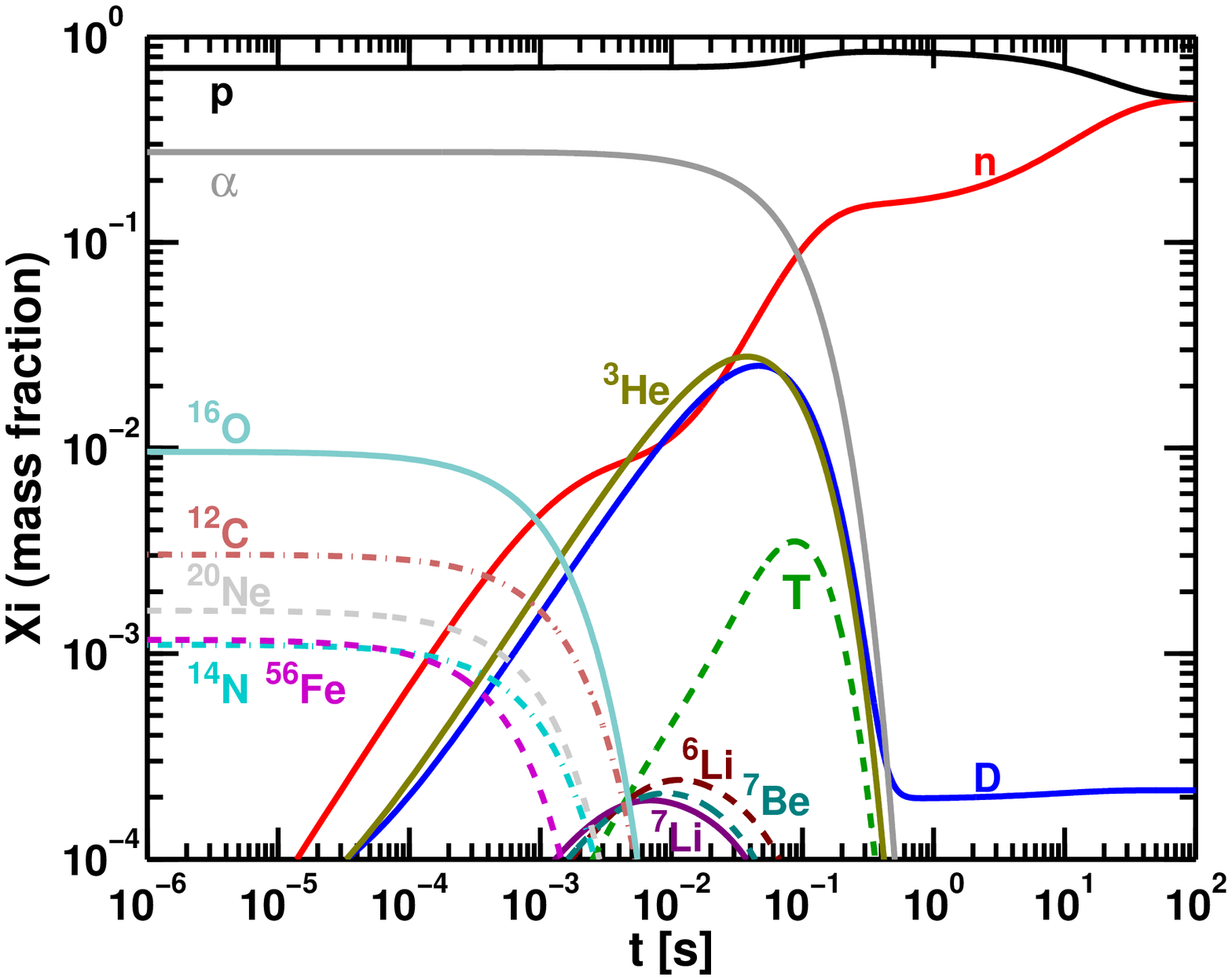}
\includegraphics[scale=0.4]{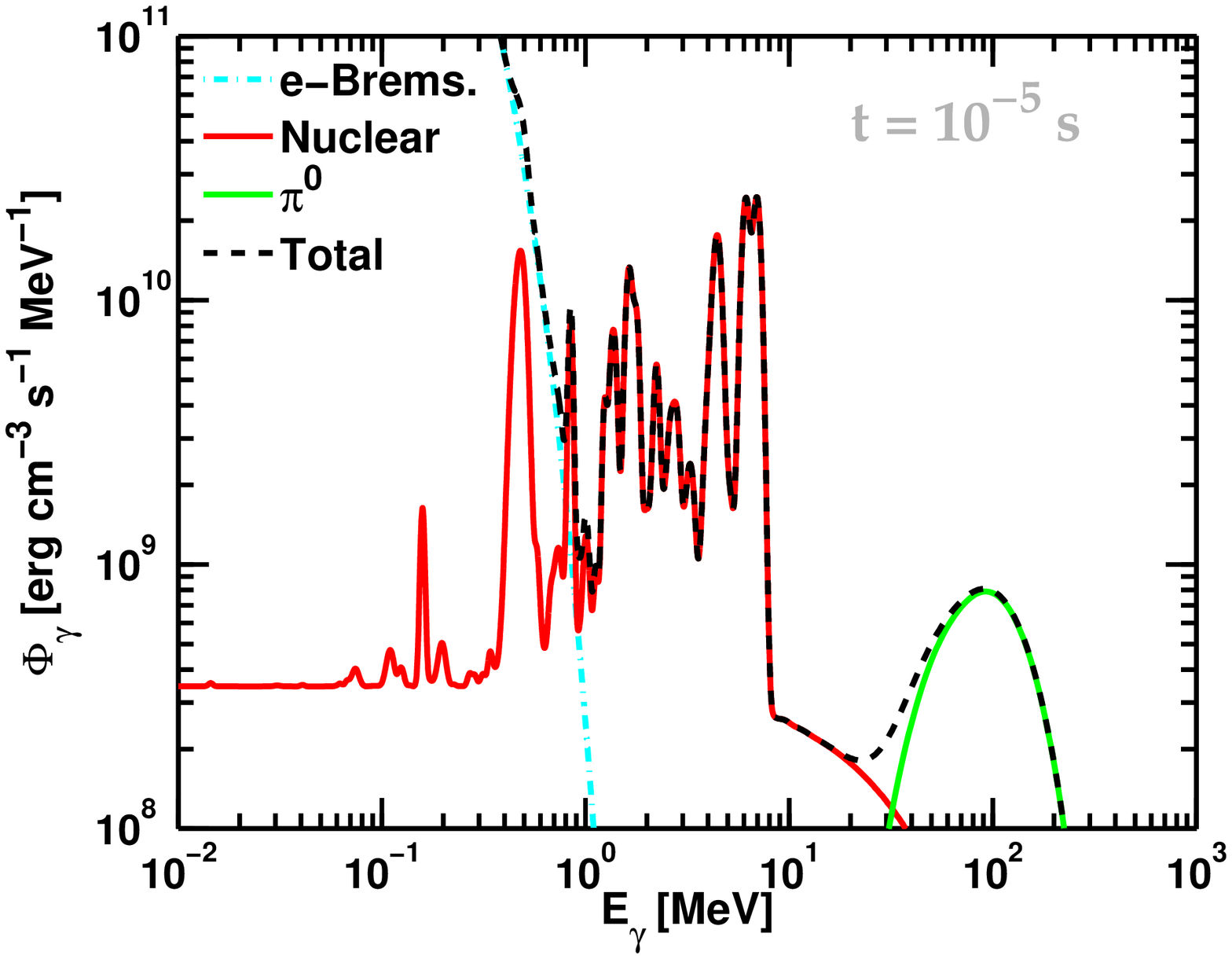}\\
\includegraphics[scale=0.4]{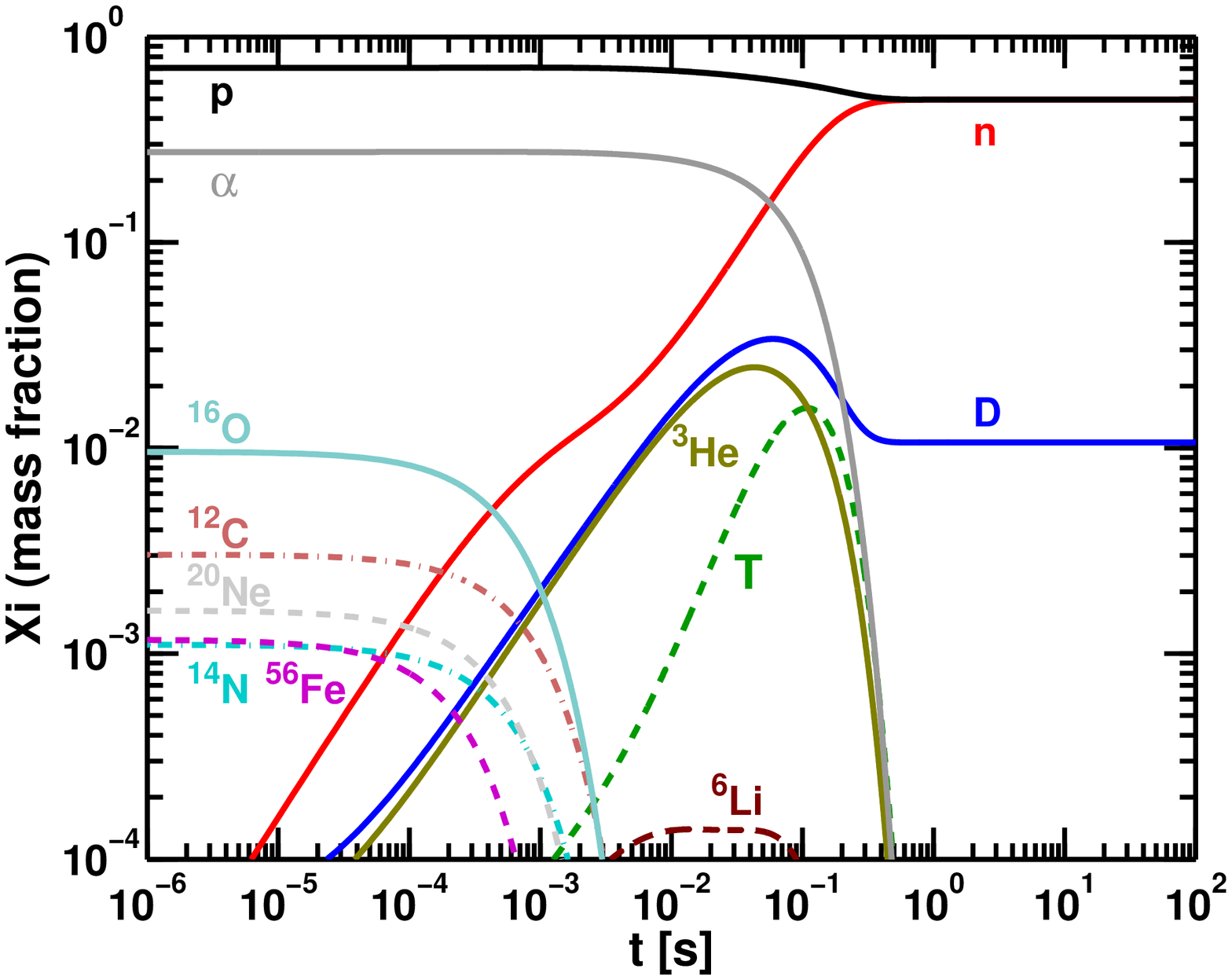}
\includegraphics[scale=0.4]{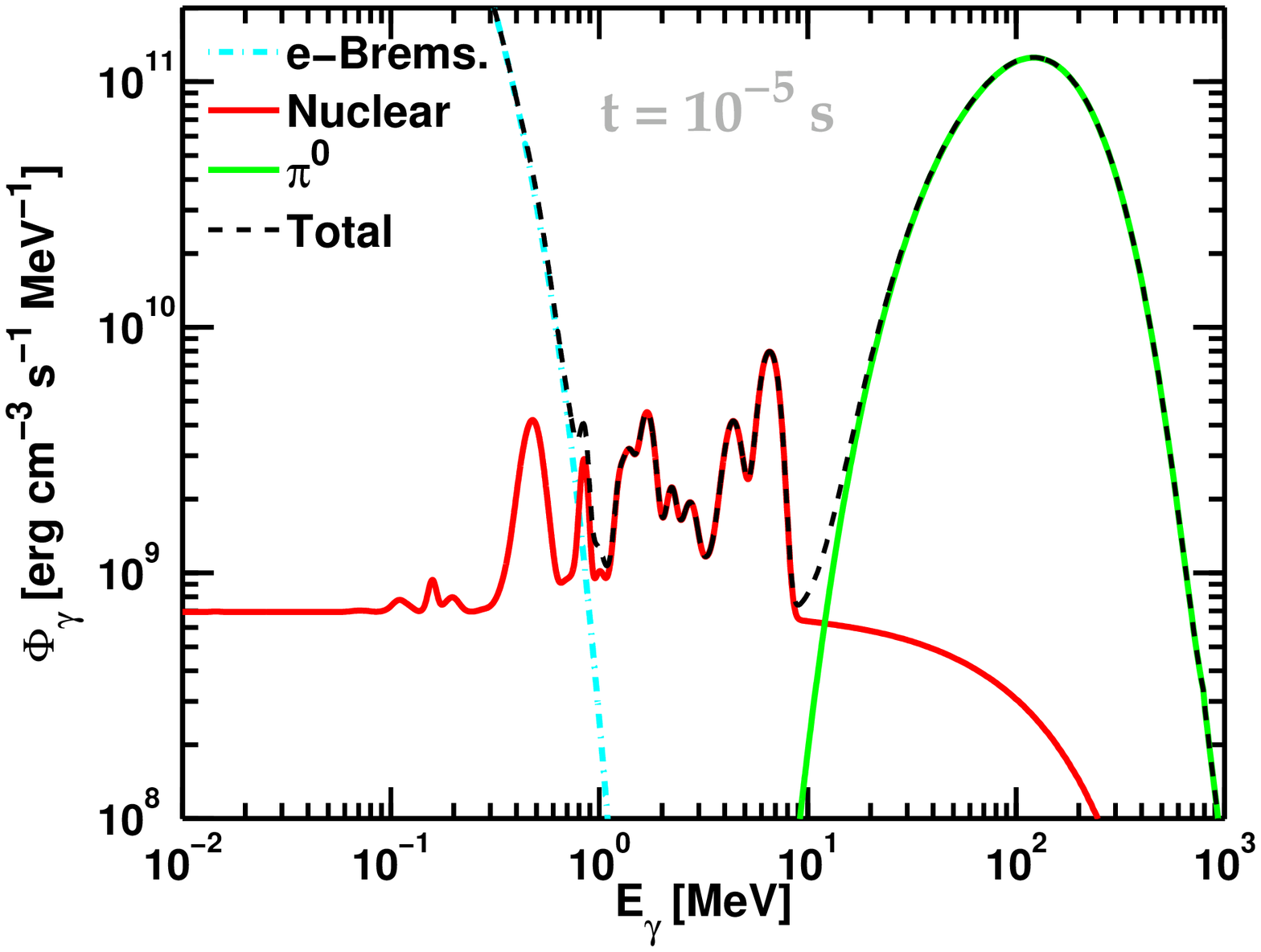}
\caption{Temporal evolution of the chemical composition of plasma 
plasma and its $\gamma$-ray emissivities at the elarly stage of the evolution, $t=10^{-5}$sec, calculated for the ion temperatures $k T_{\rm i}=30$~MeV (top panels) and $k T_{\rm i}=100$~MeV (bottom panels). The nucleon number density $\rho_{17}=1$. Panels on the left show the temporal chemical evolution of some of the important elements. Panels on the right hand side show the respective $\gamma$-ray spectrum that includes: the thermal electron bremsstrahlung at $k T_{\rm e}=100$~keV (cyan dash-dot-line), nuclear reactions emission that includes prompt $\gamma$-ray lines and continuum (red line). The green line shows the emissivity of $\pi^0$-decay $\gamma$-rays, including the contribution from nuclei. The black dashed-line is the sum of all contributions. \label{fig:nucgamspec1}}
\end{figure*}

\section{Radiation in the MeV band}

Gamma-ray emission of a very hot thermal plasma with ion temperature $T_{\rm i} >10^{10}$~K is dominated by the prompt nuclear $\gamma$-ray lines between 0.1--10~MeV and $\pi^0\to2\gamma$ decay above $10$~MeV. Unlike the $\pi^0$-meson emission, the production of nuclear $\gamma$-ray lines is sensitive to small changes of nuclear abundances. Therefore, computation of their emissivities require detailed calculations of the plasma chemical evolution. Very hot astrophysical plasmas formed in e.g. accretion disks or supernova shock waves, have low densities in comparison with e.g. central core of a star, and as a result, only the binary nuclear collisions play a role in the plasma evolution. For a binary reaction $i+j\to l +...$, the production rate per unit volume of the nucleus $l$ is \citep[see e.g.][]{Fowler1967}:
\begin{equation} \label{eq:prodRate}
\dot{n}_l = n_i\,n_j\,\left\langle \sigma v\right\rangle_{ij}^l.
\end{equation}

\noindent Here $n_i$, $n_j$ and $n_l$ are the number densities of species $i$, $j$ and $l$, respectively. The $\left\langle \sigma v\right\rangle_{ij}^l$ is the thermal reaction rate which is computed as follows:
\begin{equation}\label{eq:ReacRate}
\begin{split}
\left\langle \sigma v\right\rangle_{ij}^l(T_{\rm i}) =& (1+\delta_{ij})^{-1}\,\sqrt{\frac{8}{\pi\,\mu_{ij}\,(k T_{\rm i})^3}} \times \\
&\times \int\limits_0^\infty \sigma_{ij}^l(E)\,E\,\exp\left(-\frac{E}{k T_{\rm i}}\right)\, dE,
\end{split}
\end{equation}

\noindent where, $\delta_{ij}$ is the Kronecker delta function; $\delta_{ij}=1$ if nucleus $i$ and $j$ are identical, and zero otherwise. $\mu_{ij}=m_i\,m_j/(m_i+m_j)$ is the reduced mass of the two incoming nuclei. $\sigma_{ij}^l$ is the $i+j\to l+...$ reaction cross section. $k T_{\rm i}$ is the plasma ion temperature in units of energy and $E$ is the collision energy.

It is convenient to solve Eq.~(\ref{eq:prodRate}) in terms of the abundances (mass fraction) that for an element $i$ is defined as $X_i =A_i\,n_i/\rho $. Here, $A_i$ and $n_i$ are the mass number and the number density of an the element $i$, respectively. $\rho=\sum A_i\, n_i$ is the nucleon number density of the plasma. Utilizing $X_i$, the Eq.~(\ref{eq:prodRate}) transforms as follows:
\begin{equation}
\dot{X}^l = \rho\, \sum\limits_{i,j}^N \left(\frac{A_l}{A_i\,A_j}\right)\,\left\langle \sigma v\right\rangle_{ij}^l\, X^i\,X^j.
\end{equation}
If the plasma is composed of $N$ nuclear species, one has to sum up the contribution of all their most important reactions channels that can produce nucleus $l$. Moreover, nuclear reactions can produce nuclei that were not initially present in the plasma. Therefore, one has to consider all possible nuclear species that can be produced during the plasma lifetime and all their most important nuclear reaction channels. This system forms a so-called \textit{nuclear reaction network}. The most important quantity of the nuclear network is the production rate $\left\langle \sigma\, v\right\rangle_{ij}^l$ as a function of the plasma temperature. From Eq.~(\ref{eq:ReacRate}), calculation of the $\left\langle \sigma\, v\right\rangle_{ij}^l$ requires the reaction cross section $\sigma_{ij}^l$ that we obtain either directly from the experimental data or calculate it using the modern nuclear code TALYS \citep[][]{talys} that provides a complete description of all reaction channels and uses many state-of-the-art nuclear models, for more details see \cite{kafexhiu2012}. We note that the recent version of the code TALYS can calculate cross sections for energies up to 1~GeV. This has allowed us to extend the computation of the reaction rates $\left\langle \sigma\, v\right\rangle_{ij}^l$ for temperatures of the order $10^{12}$~K. In addition, our new nuclear reaction network considers the $p+p\to p+n+\pi^+$ and $p+p\to D+\pi^+$ channels that can significantly change the abundances of neutrons and deuterium in case of a proton (light composition) plasma with temperatures $k T_{\rm i}>10$~MeV. The cross sections for these channels are obtained by fitting the experimental data compiled in Ref.~\cite{Machner1999}.

In addition, our nuclear reaction network considers reactions $i+j\to l^*+...$ that produce nucleus $l$ in different excited states that subsequently lead to prompt $\gamma$-ray lines emission. The number of $\gamma$-rays produced per unit volume $\dot{n}_\gamma$ for a specific $l^*$ transition, is calculated using Eq.~(\ref{eq:prodRate}). Since the plasma temperature is $kT>1$~MeV, the  Doppler broadening due to thermal motion of $l^*$ is larger than the natural width of the prompt $\gamma$-ray line. Therefore, the $\gamma$-ray line profile is well described by a Gaussian and its emissivity per unit volume is given by:
\begin{equation}
\Phi_\gamma(E_\gamma) = \frac{E_\gamma^0\,\dot{n}_\gamma}{\sqrt{2\pi\,\sigma_G^2}}\,\exp\left[-\frac{(E_\gamma-E_\gamma^0)^2}{2\,\sigma_G^2}\right].
\end{equation}

\noindent Here, $E_\gamma^0$ is the central $\gamma$-ray energy obtained by the energy difference of the two transition levels of the nucleus. $\sigma_G=E_\gamma^0\,\sqrt{k T_{\rm i}/m_lc^2}$ is the Gaussian broadening and $m_l$ is the mass of the excited nucleus $l^*$. Note that in our calculations we assume that $\gamma$-rays are emitted isotropically in the rest frame of the excited nucleus. Strictly speaking, this is not the case. The accelerator measurements (see e.g. Ref.~\cite{Belhout2007, Benhabiles-Mezhoud2011}) show significant emission anisotropy which can have a non-negligible impact on the spectral distribution of radiation. Namely, it leads to the line-splitting as it has been demonstrated in Ref.~\cite{Bykov1996, Kozlovsky1997}. Calculation of this effect requires information about the $\gamma$-ray emission angular distribution for the most important nuclei and their numerous transitions. The fine structure calculation of the $\gamma$-ray line emission, is outside the scope of this paper.

In addition to nuclear lines, nuclear collisions produce a $\gamma$-ray continuum, too. This continuum emission is dominated mainly by the production and decay of the $\pi^0$-mesons. Between the nuclear lines and the $\pi^0\to2\gamma$ emission there is however a small gap that is filled by many competing channels. One of the most important channels is the nuclear bremsstrahlung which is generally dominated by the proton--nucleus emission due to higher abundance of protons. We consider this channel in our calculations, for completeness, and we approximate its spectral shape with that of the neutron--proton thermal bremsstrahlung where the normalization factor is given by $(A_t-Z_t)\,A_t^{-1/3}$ empirical law \cite{Edgington1966}. Here $Z_t$ and $A_t$ are the charge and mass numbers of the target nucleus. If the plasma lifetime is longer then the nuclear destruction timescale, a neutron--proton plasma is formed. Although, this plasma does not emit nuclear $\gamma$-ray lines, it however emits a continuum due to $n+p\to D+\gamma$ thermal capture, $n-p$ bremsstrahlung and $\pi^0\to2\gamma$ emission. Calculation of the neutron--proton capture and bremsstrahlung emission are given in Ref.~\cite{Aharonian1984}, whereas, the $\pi^0\to2\gamma$ emission is computed in the next section.

Now that we have described our nuclear reaction network, we demonstrate its solution for three specific cases. In all these examples, we keep fixed the plasma nucleon number density to $\rho_{17}=\rho/10^{17}~{\rm cm^{-3}}=1$ and the plasma temperature to $kT_{\rm i}=30$, 50 and 100~MeV, over the entire plasma evolution. The initial chemical composition was set to a solar composition. The results for the first 100~sec of the plasma evolution are shown in Figs~\ref{fig:nucgamspec} and \ref{fig:nucgamspec1}. The choice of temperature in these examples reflects the relative $\gamma$-ray emissivities between the nuclear lines and the $\pi^0$-meson production. We find that for plasma temperature around 50~MeV the $\pi^0\to2\gamma$ emissivity is comparable with that of nuclear lines. At lower temperatures, $kT_{\rm i}=30$~MeV nuclear $\gamma$-ray lines dominate and at higher temperatures, $kT_{\rm i}=100$~MeV the $\pi^0\to2\gamma$ emissivity dominates. Note that the results presented in Figs~\ref{fig:nucgamspec} and \ref{fig:nucgamspec1} can be scaled for an arbitrary density simply by rescaling the evolution time $t \sim \rho^{-1}$ and the plasma emissivity $\Phi_\gamma \sim\rho^2$.

\section{$\pi^0$-meson production}

\subsection{Production rates \label{sec:Formalism}}
Unlike the nuclear reactions, $\pi^0$-meson production is sensitive on the high energy tail of the plasma distribution function. Therefore, we consider the relativistic formulation of the plasma quantities. For a relativistic and isotropic plasma,  Eq.~(\ref{eq:ReacRate}) transforms to (see e.g. Ref.~\cite{weaver1976}):
\begin{equation}
\label{eq:drate}
\left\langle \sigma v\right\rangle_{ij}^l =\int \frac{f_i(\gamma_i)f_j(\gamma_j)}{(1+\delta_{ij})} \, \sigma_{ij}^l(\gamma_r)\frac{\sqrt{\gamma_r^2 -1}}{\gamma_i\gamma_j} \, d\gamma_id\gamma_j \frac{du}{2}.
\end{equation}
Here, $\gamma_i=E_i/m_i$ and $\gamma_j=E_j/m_j$ are the particles $i$ and $j$ Lorentz factors in the laboratory frame (LAB), respectively; $E_i$, $E_j$, $m_i$ and $m_j$ are particles energies and masses; $u=\cos(\eta)$, where $\eta$ is the angle between the two colliding particles in the LAB frame; $f_i(\gamma_i)$ and $f_j(\gamma_j)$ are the energy distribution functions for particle $i$ and $j$ in the plasma, respectively; $\gamma_r=\gamma_i\gamma_j(1-\beta_i\beta_j\,u)$ is the collision Lorentz factor, where $\beta_i$ and $\beta_j$ are the particles speeds (in units of $c$), and $c$ is the speed of light in the vacuum.

If the species $i$ and $j$ are fully thermalized in the plasma, their energy distribution functions are given by the relativistic version of the Maxwell distribution, the so-called Maxwell-J\"uttner distribution \citep[see e.g.][]{Juttner1911,Juttner1911b,MaxwellJuttner2010},
\begin{equation}
f_{MB}(\gamma,\theta) = \frac{\gamma^2\beta}{\theta\,K_2(1/\theta)}\exp\left(-\frac{\gamma}{\theta} \right),
\end{equation}
\noindent where, $\gamma$ and $\beta$ are the particle Lorentz factor and speed, respectively; whereas, $\theta=k T_{\rm i}/m\,c^2$ is the dimensionless temperature; $T_{\rm i}$ is the plasma temperature and $m$ is the particle mass. $K_2(x)$ is the modified Bessel function of the second kind.

Because of the plasma short lifetime and/or instabilities, the Maxwellian high energy tail may not be a good representation of the distribution function; it could vary from a sharp cutoff to a power-law function. We model these variations with two types of distribution functions: a Maxwellian distribution with a sharp cutoff ($f_{\rm{cut}}$) and a Maxwellian distribution with a power-law tail at high energies ($f_{\rm{pl}}$). They are defined as follows:
\begin{equation} \label{eq:distcut}
f_{\rm{cut}}(\gamma,\tilde{\theta})~~= A\left\{
   \begin{array}{ll}
    f_{MB}(\gamma,\tilde{\theta}) & : 1\leq\gamma\leq\gamma_0\\
    0 & :\gamma>\gamma_0
   \end{array}\right. \\
\end{equation}

\begin{equation}\label{eq:distpwl}
f_{\rm{pl}}(\gamma,\tilde{\theta},\alpha) = A\left\{
   \begin{array}{ll}
    f_{MB}(\gamma,\tilde{\theta}) & : 1\leq\gamma\leq\gamma_0\\
    B\times E_{\rm k}^{-\alpha} & :\gamma>\gamma_0
   \end{array}
  \right.
\end{equation}
Here, $\gamma$ is the Lorentz factor of the particle; $\tilde{\theta}$ is the dimensionless temperature that is estimated from the particles average energy for the specific distribution; $E_{\rm k}$ is the kinetic energy per nucleon and $\alpha$ is the power-law index related to the acceleration mechanism. $A$ is the normalization of the distribution function, whereas, the constant $B$ is fixed to ensure the continuity of the distribution function at $\gamma_0$. 

For comparison of calculations using the pure and the modified Maxwellian distributions, the total number of particles and the energy must be conserved. For this reason we start with a pure Maxwellian distribution with a given temperature $\theta$ and then modify it to $f_{\rm{cut}}$ or $f_{\rm{pl}}$. We fix the constants $A$ and $\tilde{\theta}$ from the conservation condition of the total number of particles and the plasma energy: 
\begin{equation}
\label{eq:distEcons}
\left\{
\begin{array}{l}
\int\limits_1^\infty\,f(\gamma,\tilde{\theta})\,d\gamma = 1\\
\int\limits_1^\infty \,(\gamma-1)\,f(\gamma,\tilde{\theta})\,d\gamma = \int\limits_1^\infty\,(\gamma-1)\,f_{MB}(\gamma,\theta)\,d\gamma
\end{array}\right.
\end{equation}

\subsection{Maxwellian distribution}

\subsubsection{$p+p\to\pi^0$ production rate}
\begin{figure}
\includegraphics[scale=0.45]{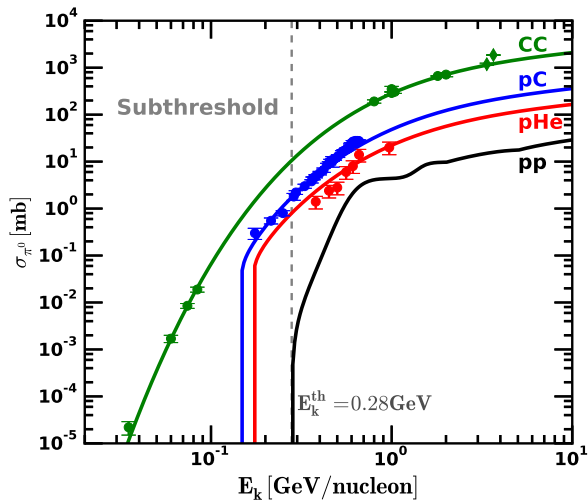}
\caption{$\pi^0$-meson production cross section as a function of the projectile kinetic energy per nucleon for reactions $p+p\to\pi^0$ (black), ${\rm p+ ^4He \to\pi^0}$ (red), ${\rm p+ ^{12}C \to\pi^0}$ (blue) and ${\rm ^{12}C+ ^{12}C \to\pi^0}$ (green). The experimental data points are from Refs.~\cite{Bayukov1957, Pollack1963, Dunaitsev(1964), Aksinenko1980, Bellini1989, Holzmann(1997)pi01GeV, Averbeck2003, Laue2000, Agakishiyev(1985)pim}, whereas, the lines represent their analytical parametrisations given in Ref.~\cite{kafexhiu2014, Subthresh2016}.\label{fig:XSsubPi0}}
\end{figure}

Generally, the hydrogen is the most abundant element everywhere, hence it should dominate also in astrophysical plasmas. Consequently, the $p+p\to\pi^0$ reaction is the most important channel of $\pi^0$-meson production. The pion production rate in thermal plasma can be calculated using the formalism of Ref.~\cite{weaver1976}. For the plasma temperatures less than $10^{12}$K, the calculations are most sensitive to the $p+p\to\pi^0$ cross-section close to the kinematic threshold of $\pi^0$-meson production. An accurate parametrization of the cross section down to the kinematic threshold at 280~MeV, recently has been derived in Ref.~\cite{kafexhiu2014}, using the most relevant publicly available experimental data. The black line in Fig~\ref{fig:XSsubPi0} shows the inclusive $p+p\to\pi^0$ production cross section given by this parametrization.

\begin{figure}
\includegraphics[scale=0.45]{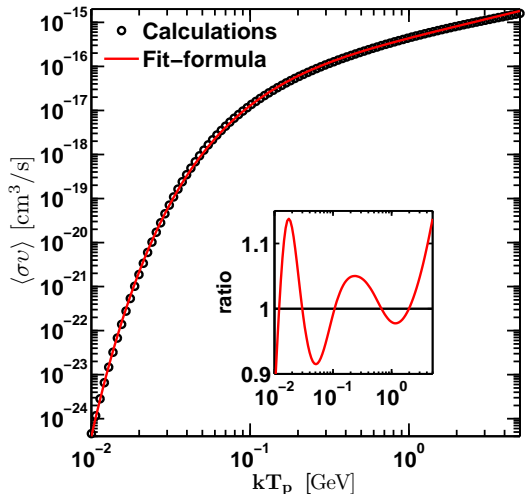}
\caption{The $p+p\to\pi^0$ thermal production rate $\left\langle \sigma\,v\right\rangle_{pp}^{\pi^0}$ as a function of proton plasma temperature. The open circles represent the computation result using Eq.~(\ref{eq:drate}) relativistic formula \citet{weaver1976}. The red line is the fitting formula shown in Eq.~(\ref{eq:ppRFit}). The accuracy of the parametrization is better than 15\% for $10~{\rm MeV}\leq k T_{\rm p} \leq 5$~GeV as is shown in the inserted figure. 
\label{fig:ppRFit}}
\end{figure}

Using Eqs.~(\ref{eq:drate}) and the $\sigma_\pi$ formulas described in Ref.~\cite{kafexhiu2014}, we compute the thermal $\pi^0$-meson production rate; the results are shown in Fig~\ref{fig:ppRFit}, where the $\left\langle \sigma \upsilon \right\rangle_{pp}^{\pi^0}$ is plotted as a function of the proton plasma temperature. This relation can be presented by a simple analytical form:
\begin{equation}\label{eq:ppRFit}
\begin{split}
\left\langle \sigma \upsilon \right\rangle_{pp}^{\pi^0} &= (1+\theta_p)^{3/4} \times \\ & \exp\left(-0.569\,\theta_p^{-0.7865} -35.38\,\theta_p^{-0.0035} \right) {\rm \left[\frac{cm^3}{s}\right]}, 
\end{split}
\end{equation}
where $\theta_p=k T_{\rm p}/m_pc^2$ is the dimensionless proton temperature. The parametrization is valid for ion temperatures between $10~{\rm MeV}\leq k T_{\rm p} \leq 5$~GeV and provides an accuracy better than 15~\% in the entire range of temperatures (see Fig~\ref{fig:ppRFit}).

\subsubsection{Emissivity}

For computing the $\gamma$-ray spectrum from the $\pi^0\to 2\gamma$ decay channel, the $\pi^0$-meson production invariant differential cross section is required. In plasma, the colliding protons can have arbitrary speeds and therefore, computation of the spectra of $\pi^0$-mesons involve many kinematic transformations. To simplify the calculations, we propose an analytical representation of the $p+p\to \pi^0 \to 2\gamma$ differential cross section at low collision energies near the kinematic threshold for arbitrary speeds of the two colliding protons (see Appendix~\ref{sec:Appendix}). We have constructed the neutral pion invariant differential cross section $E_\pi\times d\sigma/dp_\pi^3$ in the center-of-mass reference frame assuming isotropic emission of pions at low energies and, adopting the center of mass parametrization of the experimental $p+p\to\pi^0$ energy distributions \cite{kafexhiu2014}. Finally, we performed the relevant kinematic transformations for the $\pi^0\to2\gamma$ decay for the plasma rest frame of reference. 

\begin{figure}
\includegraphics[scale=0.235]{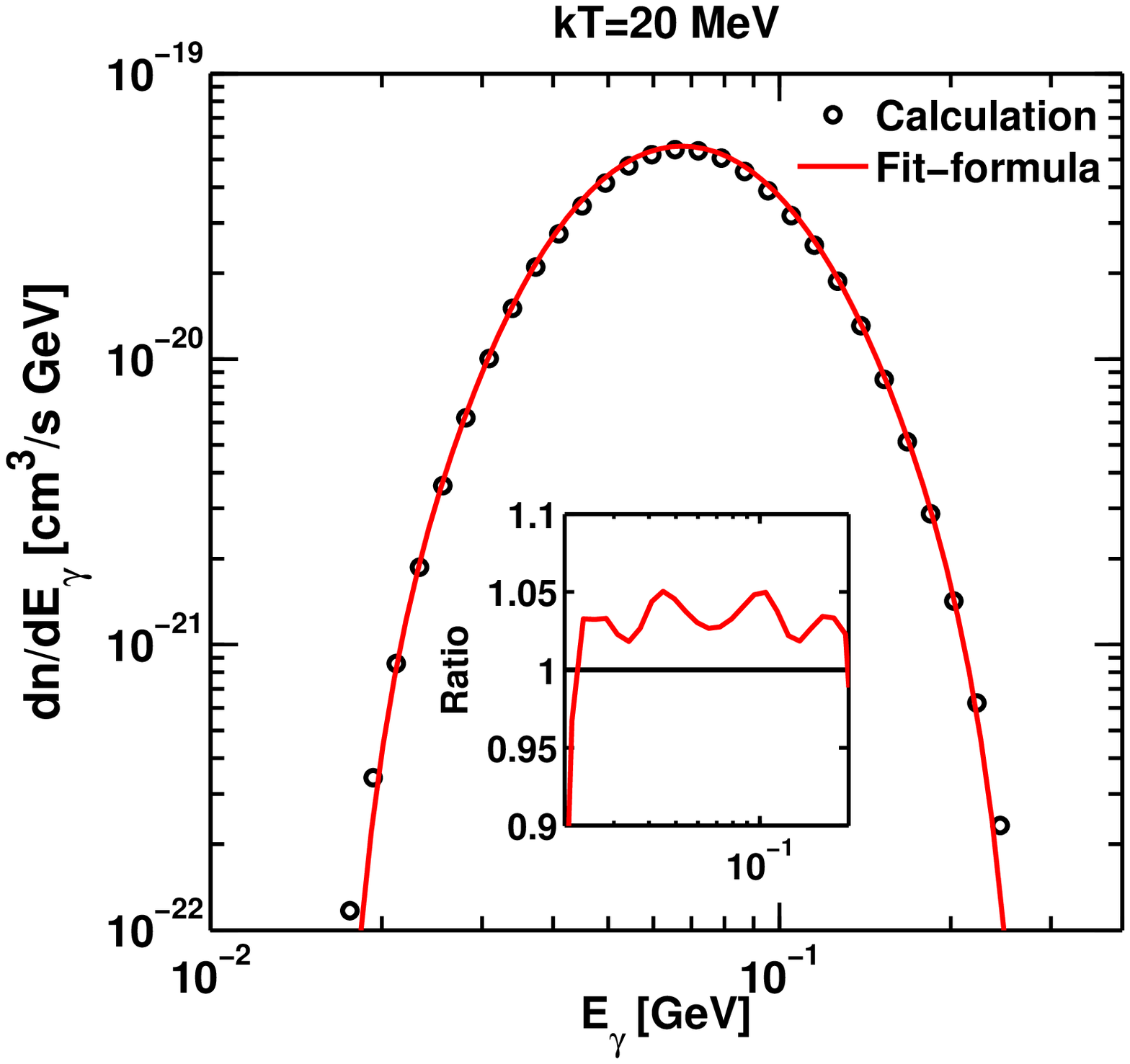}
\includegraphics[scale=0.235]{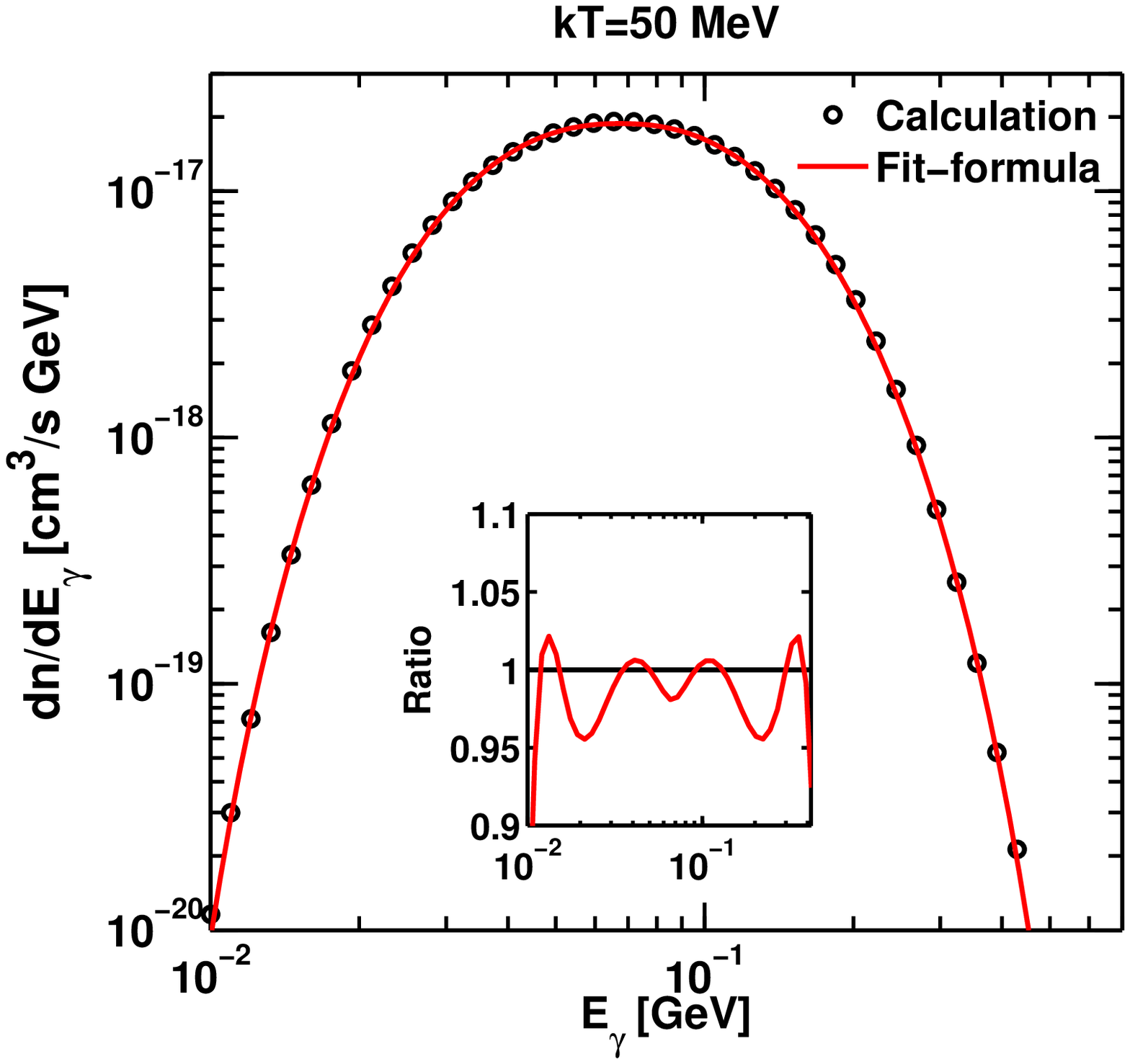}\\
\includegraphics[scale=0.235]{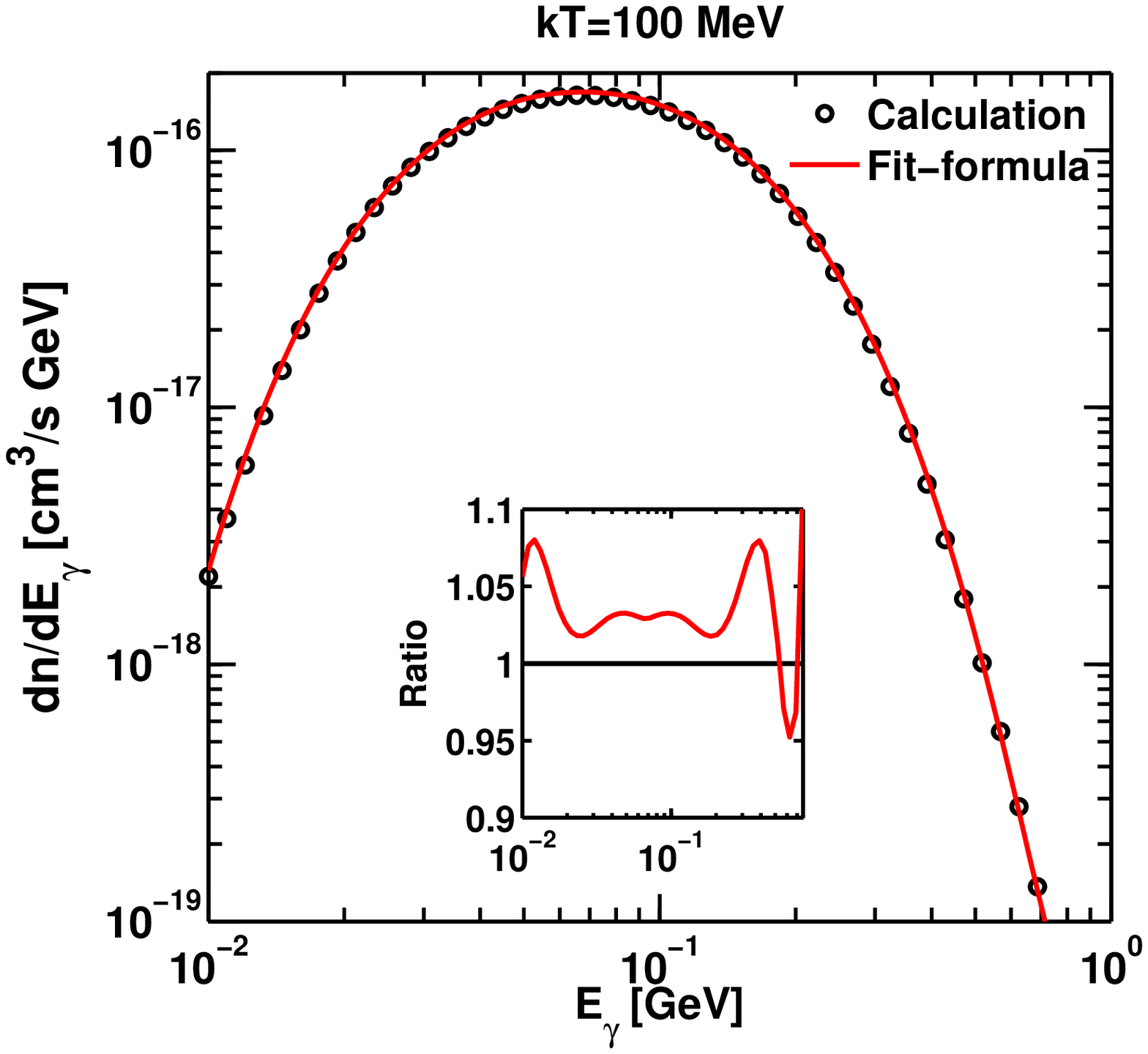}
\includegraphics[scale=0.235]{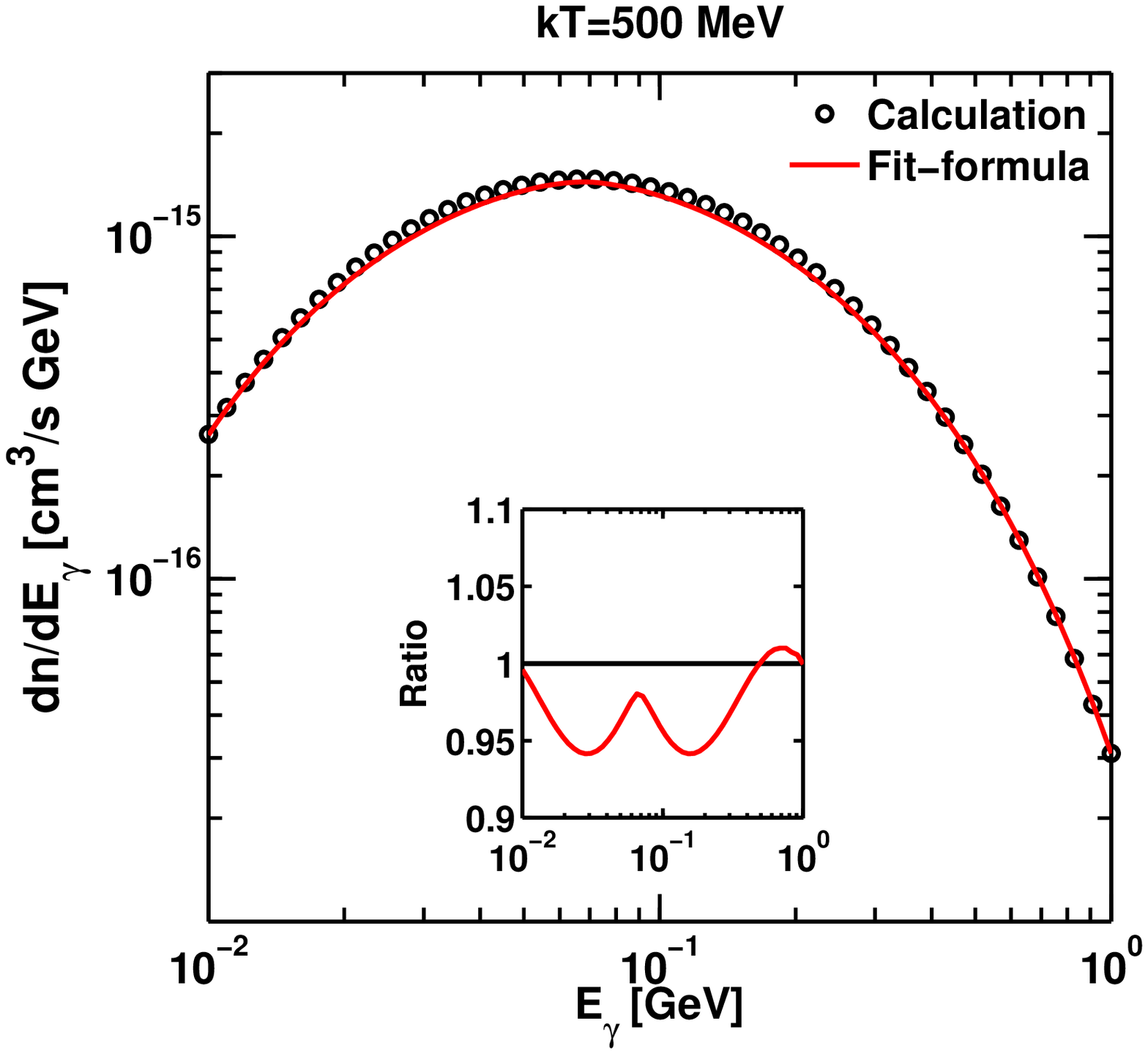}
\caption{The $\pi^0$-decay $\gamma$-ray emissivity of the Maxwellian plasma for four proton plasma temperatures $k T_{\rm p}=20$, 50, 100 and 500 MeV. Black open circles represent the numerical calculations using the parametrization of the differential cross section given in the Appendix~\ref{sec:Appendix}. The red line represents the parametrization given in Eq.~(\ref{eq:emisivity}). The inserted figures show the ratio between the results of the parametrization and the numerical calculations. 
\label{fig:emis}}
\end{figure}

The $\gamma$-ray emissivities of plasma for a Maxwellian distribution of protons calculated for different temperatures, are shown in Fig.~\ref{fig:emis}. To make these results more accessible, we have parametrized them analytically as a function of the proton temperature. We present the differential $\gamma$-ray emissivity, $d\dot{n}_\gamma/dE_\gamma$ -- the $\gamma$-ray production rate per unit volume and per unit energy interval -- as a product of two functions,
\begin{equation}\label{eq:emisivity}
\frac{d\dot{n}_\gamma}{dE_\gamma}(E_\gamma,\theta_p) = n_p^2\times\mathcal{S}_{\rm max}(\theta_p) \times F\left(\theta_p,E_\gamma\right).
\end{equation}
The function $F\left(\theta_p,E_\gamma\right)$ determines the spectral shape of $\gamma$-rays for the given temperature $\theta_p$; it varies between 0 and 1. The second function $\mathcal{S}_{\rm max}$ corresponds to the maximum of the emissivity which is obtained at the energy $E_\gamma=m_{\pi^0}c^2/2\approx67$~MeV. It depends only on the proton temperature:
\begin{equation}
\mathcal{S}_{\rm max}(\theta_p) = \exp\left(\frac{1.224\,x^3 +7.18\,x^2 + 29.17\,x -11.29}{\sqrt{-x}} \right),
\end{equation}
where $x=\log(\theta_p)$. This function is in units ${\rm cm^{-3}s^{-1} GeV^{-1}}$ and is valid for $20\leq k T_{\rm p} \leq 500$~MeV. For the function $F$, we offer the following form:
\begin{equation}
F\left(\theta_p,E_\gamma\right) = \exp\left(\alpha(\theta_p)\,X_\gamma^{\beta(\theta_p)} + \delta(\theta_p)\,X_\gamma^6\right), 
\end{equation}
where 
\begin{equation}
X_\gamma=\frac{Y_\gamma-m_{\pi^0}c^2}{Y_\gamma^0-m_{\pi^0}c^2}, 
\end{equation}
and 
\begin{equation}
Y_\gamma = E_\gamma + \frac{m_{\pi^0}^2 c^4}{4\,E_\gamma}, 
Y_\gamma^0=E_\gamma^0+ \frac{m_{\pi^0}^2 c^4}{4\,E_\gamma^0}, 
\end{equation}
where $E_\gamma^0=7\, k T_{\rm p} + m_{\pi^0}c^2/2$. This formula is valid only for $E_\gamma \leq E_\gamma^0$. 

For the interval of temperatures $20\leq k T_{\rm p} \leq 150~\rm{MeV}$, 
$\alpha(\theta_p)$, $\beta(\theta_p)$ and $\delta(\theta_p)$ are parametrized as:
\begin{equation}
\begin{split}
\alpha(x)&=1.23\,x^3 +11.09\,x^2 +29.5\,x +15.68 \\
\beta(x) &= -0.161\,x^2 -0.955\,x -0.281\\
\delta(x) &= -1.048\,x^4 -12.826\,x^3 -57.514\,x^2 -\\
  & \hspace{2.4cm} -111.287\,x -77.8~.
\end{split}
\end{equation}
For the temperatures between $150\leq k T_{\rm p} \leq 500~\rm{MeV}$, they become:
\begin{equation}
\begin{split}
\alpha(x)&= -0.49\,x^2 -4.09\,x -14.6 \\
\beta(x) &= -0.083\,x + 0.79 \\
\delta(x) &= 3.145\,x^3 + 13.183\,x^2 + 17.47\,x + 7.2~.
\end{split}
\end{equation}

The parametrization given in Eq.~(\ref{eq:emisivity}) provides an accuracy better than 10~\% for the entire temperature range $20\leq k T_{\rm p} \leq 500$~MeV and $E_\gamma \leq E_\gamma^0$ (see Fig.~\ref{fig:emis}).

\subsubsection{Contributions from heavier nuclei}

\begin{figure}
\includegraphics[scale=0.45]{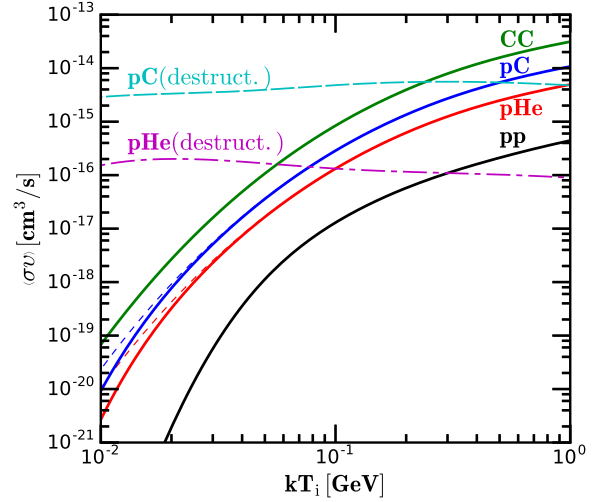}
\caption{Neutral pion production rate in the Maxwellian plasma for different reactions, $\left\langle \sigma\,v\right\rangle_{ij}^{\pi^0}$ as a function of the plasma temperature. The short-dash lines represent the numerical calculations and the solid lines their respective production rates using the parametrization given by Eq.~(\ref{eq:subRate}). The cyan long-dash line and the magenta long-dash-dot line are the Carbon and Helium destruction rates, respectively. \label{fig:ratesub}}
\end{figure}

\begin{figure}
\includegraphics[scale=0.45]{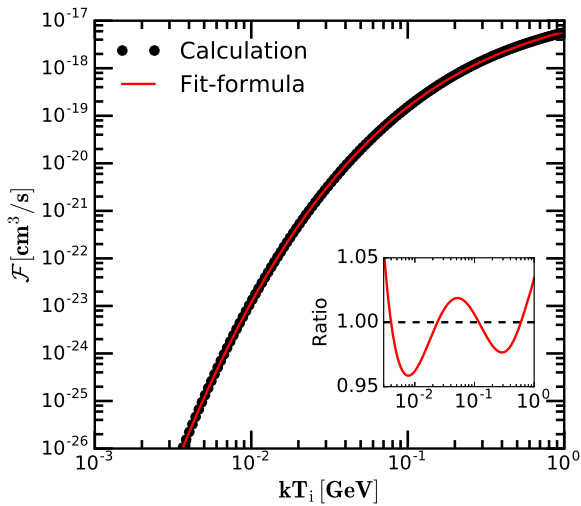}
\caption{$\mathcal{F}$ function described in Eq.~(\ref{eq:subRate}) as a function of the plasma temperature. Black open circles show the numerical calculations, whereas, the red line is its parametrization given in Eq.~(\ref{eq:subRate}). \label{fig:ratesubpart}}
\end{figure}

At ion temperatures $k T_{\rm i} \geq 10$~MeV, when the $\pi^0$ production becomes an important process, the nuclei heavier than hydrogen are rapidly destroyed in inelastic collisions. Up to $k T_{\rm i} \sim 100$~MeV, the destruction rates are orders of magnitude larger than the pion production rate (see Fig.~\ref{fig:ratesub}), therefore, their contribution can be safely ignored. Nevertheless, for a purpose of consistency, we present here the production rates of $\pi^0$-mesons from nuclei heavier than hydrogen. For these calculations we use the recent parametrizations \cite{Subthresh2016} of nucleus--nucleus cross sections including the so-called \textit{subthreshold pion} production. In Fig.~\ref{fig:ratesub} we show the $\pi^0$-meson production rates for some nucleus-nucleus interactions as a function of plasma temperature. By neglecting the differences of nucleus--nucleus $\pi^0$-meson production cross sections near the kinematic threshold, we suggest a universal function that describes with a good accuracy the $\pi^0$-meson production rates. For two colliding nuclei with mass numbers $A_i$ and $A_j$, the production rate, 
$\left\langle\sigma\upsilon \right\rangle_{ij}^{\pi^0}$ can be written in the following form
\begin{equation}\label{eq:subRate}
\begin{split}
\mathcal{F} & = \exp\left[0.0568\,x^3 -0.168\,x^2 + 0.94\,x -39.73 \right]\\
\mathcal{C}_{ij} & = \left(A_i A_j^{2/3} + A_i^{2/3} A_j\right)\left(1+\frac{2}{A_i^{1/3}+A_j^{1/3}}\right)^2\\
\left\langle\sigma\upsilon \right\rangle_{ij}^{\pi^0} & = \mathcal{C}_{ij} \times \mathcal{F}(\theta_p) ~~~~ \left[{\rm \frac{cm^3}{s}}\right].
\end{split}
\end{equation}
where, $x=\log(\theta_p)$ and $\theta_p=k T_{\rm i}/m_p c^2$. The above parametrization is valid for the temperatures $k T_{\rm i} \leq 1~\rm{GeV}$. It deviates from the numerical calculations by less than 5~\% (see Fig.~\ref{fig:ratesubpart}). 

\subsubsection{Cooling rate of plasma through the $p+p\to \pi^0$ channel}

\begin{figure}[h]
\centering
\includegraphics[scale=0.45]{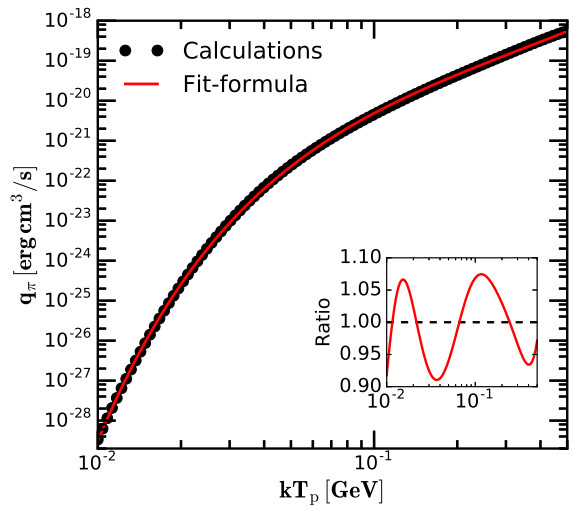}
\caption{The $p+p\to\pi^0$ cooling rate for a thermal proton plasma as a function of the plasma temperature. The open circles are calculations, whereas, the red line is the parametrization formula given in Eq.~(\ref{eq:loss}). The accuracy of this parametrization, as it can be seen from the ratio, is better than 9\% and is valid for $10\leq k T_{\rm p} \leq 500~\rm{MeV}$. \label{fig:loss}}
\end{figure}

The energy loss of a thermal proton plasma due to $\pi^0$-meson emission is calculated from $q_\pi=\int E_\pi\,f_\pi(E_\pi)\,dE_\pi$. Here, $f_\pi(E_\pi)$ is the $\pi^0$-meson emissivity spectrum. We have calculated this quantity as a function of the plasma temperature and the results are shown in Fig.~\ref{fig:loss}. In addition, we have parametrized analytically the $p+p\to\pi^0$ cooling rate as a function of the plasma temperature as follows:
\begin{equation}\label{eq:loss}
q_\pi = 10^{-3} \times \left( 1+\theta_p\right)^3 \,\theta_p^{0.57} \, \left\langle \sigma \upsilon \right\rangle_{pp}^{\pi^0} ~{\rm \left[erg~\frac{cm^3}{s}\right]},
\end{equation}

\noindent where, $\left\langle \sigma \upsilon \right\rangle_{pp}^{\pi^0}$ is given in Eq.~(\ref{eq:ppRFit}). This formula is valid for $10\leq k T_{\rm p} \leq 500~\rm{MeV}$ and has an accuracy better than 9~\% as it can be seen from the ratio between the numerical calculations and the parametrization in Fig.~\ref{fig:loss}.

\subsubsection{Cooling rates of plasma through different channels}

\begin{figure}
\includegraphics[scale=0.5]{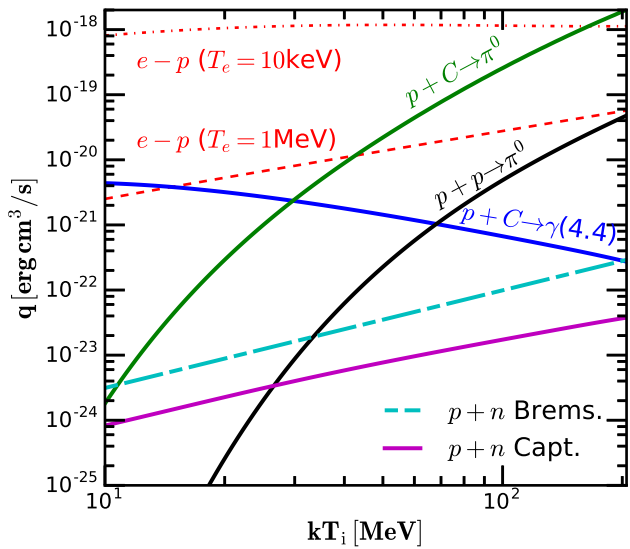}
\caption{Ion cooling rate as a function of the ion temperature for some important very hot plasma processes. The red dash-dot and dash lines represent the cooling rates from the Coulomb interaction with the electron gas of temperature $k T_e=10$~keV and 1~MeV, respectively. The black solid line is the cooling rate from the $p+p\to\pi^0$ production. The green solid line is the cooling rate for $p+^{12}$C$\to\pi^0$ and the blue solid line is the cooling rate for $p+^{12}$C$\to\gamma(4.4)$ prompt $\gamma$-ray line. For both these lines a plasma with equal number of protons and $^{12}$C was assumed. The long-dash cyan line and the solid magenta line represent the cooling rates for a neutron--proton plasma with equal number of neutrons and protons, due to the $n+p$ bremsstrahlung and the $n+p\to D+\gamma(2.22{\rm MeV})$ radiative capture, respectively. \label{fig:Qcool}}
\end{figure}

In Figure~\ref{fig:Qcool}, we show the plasma cooling rates as a function of the ion temperature for several important radiative channels. The Coulomb exchange with electrons is the most important processes of energy dissipation for ions (see e.g. Ref.~\cite{Stepney1983}). This process, operating together with the radiation of electrons (through bremsstrahlung or Comptonization), establishes the electron and ion temperatures of plasma. The cooling rate of the proton component of plasma due to the Coulomb exchange is shown in Fig.~\ref{fig:Qcool} for two electron temperatures $k T_{\rm e}=10$~keV and 1~MeV. At very high temperatures, the radiative cooling of protons proceeds through the $\pi^0$-meson production channel. 

As we saw in the previous subsection, the energy loss due to $p+p\to\pi^0$ emission increases with proton temperature. The cooling rate of a proton plasma due to pion emission (mainly $\pi^+$-mesons) can be as large as five times that of $\pi^0$-meson due to larger $p+p\to\pi^+$ cross sections.

When heavier nuclei are present in the plasma, they also loose energy through the $\gamma$-ray line emission and the continuum radiation dominated by the pion production. For plasma temperatures $k T_{\rm i}\lesssim 20$~MeV the losses due to $\gamma$-ray line emission dominates, whereas, at higher temperatures the $\pi$-meson production dominates. In Fig.~\ref{fig:Qcool} we illustrate the energy loss rates for the 4.44~MeV prompt $\gamma$-ray line emission and the $\pi^0$-meson production from the $p+^{12}$C interaction assuming that plasma is composed of an equal number of protons and carbon ions. 

If no fresh matter is supplied, very hot plasma rapidly evolves toward a neutron--proton plasma. The main cooling processes of a neutron--proton plasma are the $n+p$ bremsstrahlung and the $n+p\to D+\gamma(2.22{\rm MeV})$ radiative capture. The cooling rates for these processes are shown in Fig.~\ref{fig:Qcool} and are computed following Ref.~\citep{Aharonian1984, kafexhiu2012}. For these calculations we have assumed a plasma that has equal numbers of neutrons and protons. We find that these losses have a simple temperatures dependence for $k T_{\rm i} \gg 1$~MeV that can be approximated as follows:
\begin{equation}
\begin{split}
q_{np}^{\rm brems} &\approx 9.0\times 10^{-26}\times \theta_p^{3/2}~~[{\rm erg\,cm^3\,s^{-1}}],\\
q_{np}^{\rm capt~~} &\approx 4.6\times 10^{-26}\times \theta_p^{1.3}~~\,[{\rm erg\,cm^3\,s^{-1}}].\\
\end{split}
\end{equation}

Note that Fig.~\ref{fig:Qcool} calculations can be easily used for $n/p$ or ${\rm ^{12}C/p}$ ratios different from one by simply multiplying these results with the actual ratios.

\subsection{Modified Maxwellian distributions \label{sec:ModifTail}}

\begin{figure*}
\includegraphics[scale=0.4]{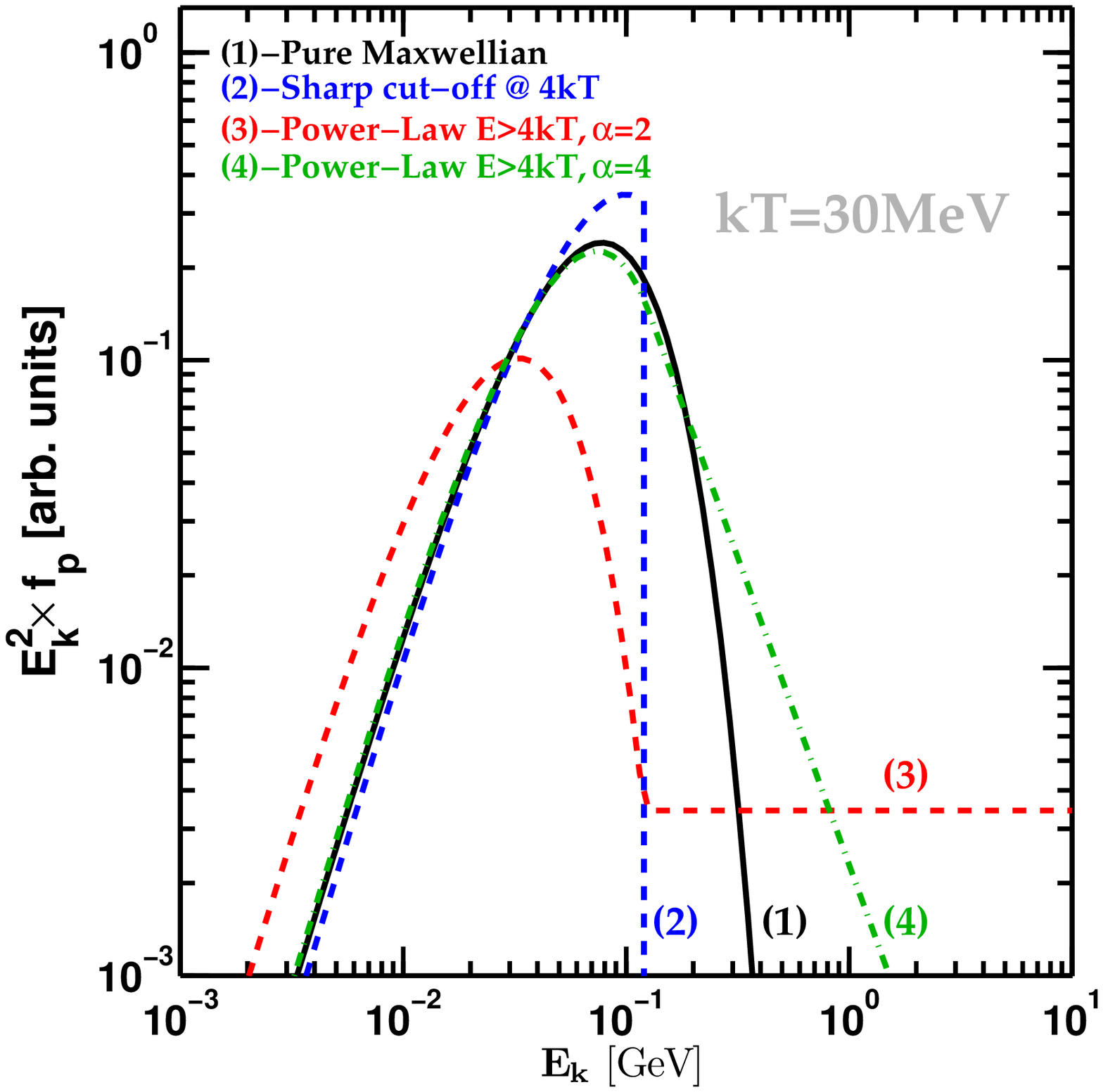}
\includegraphics[scale=0.4]{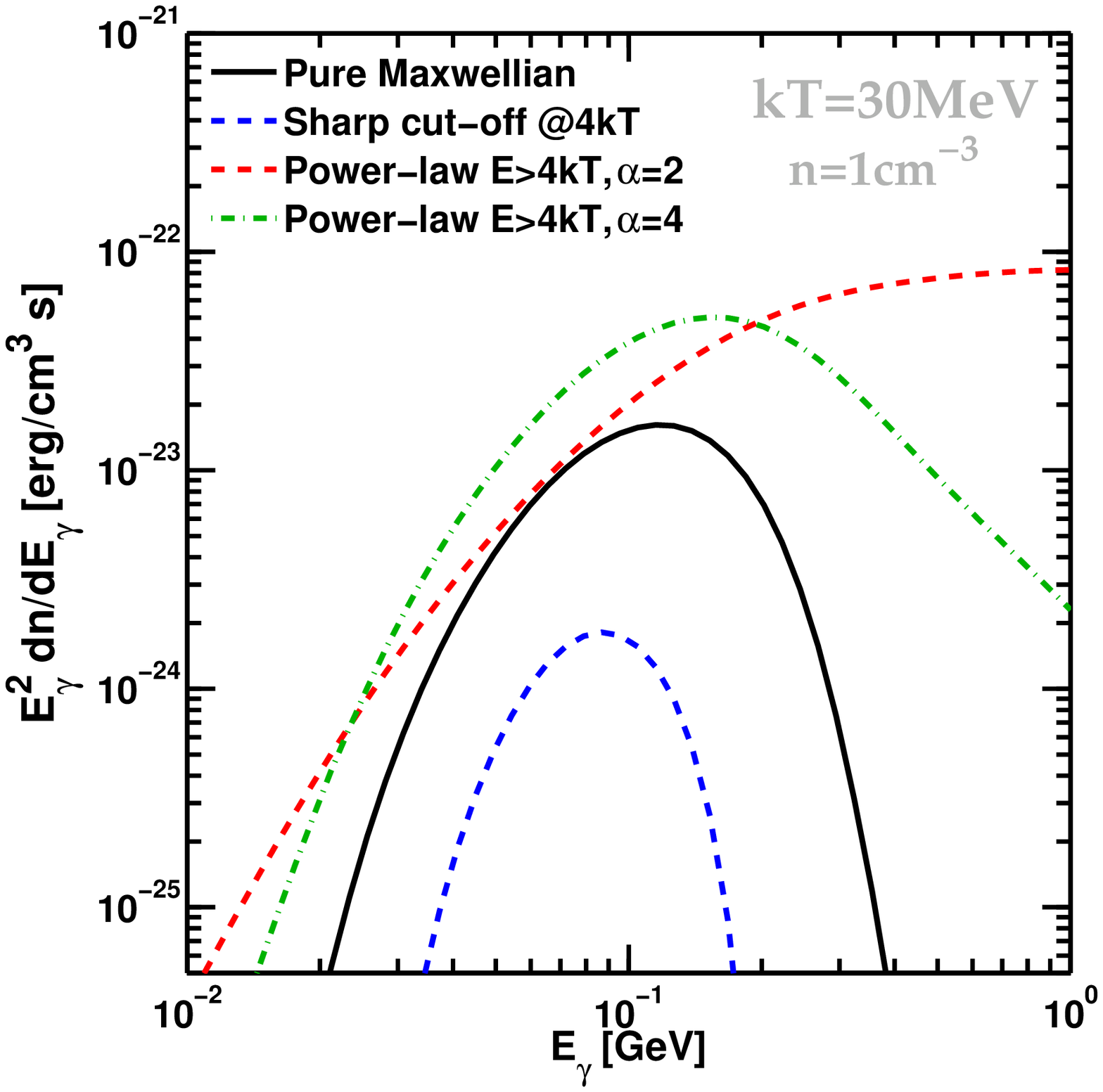}\\
\includegraphics[scale=0.4]{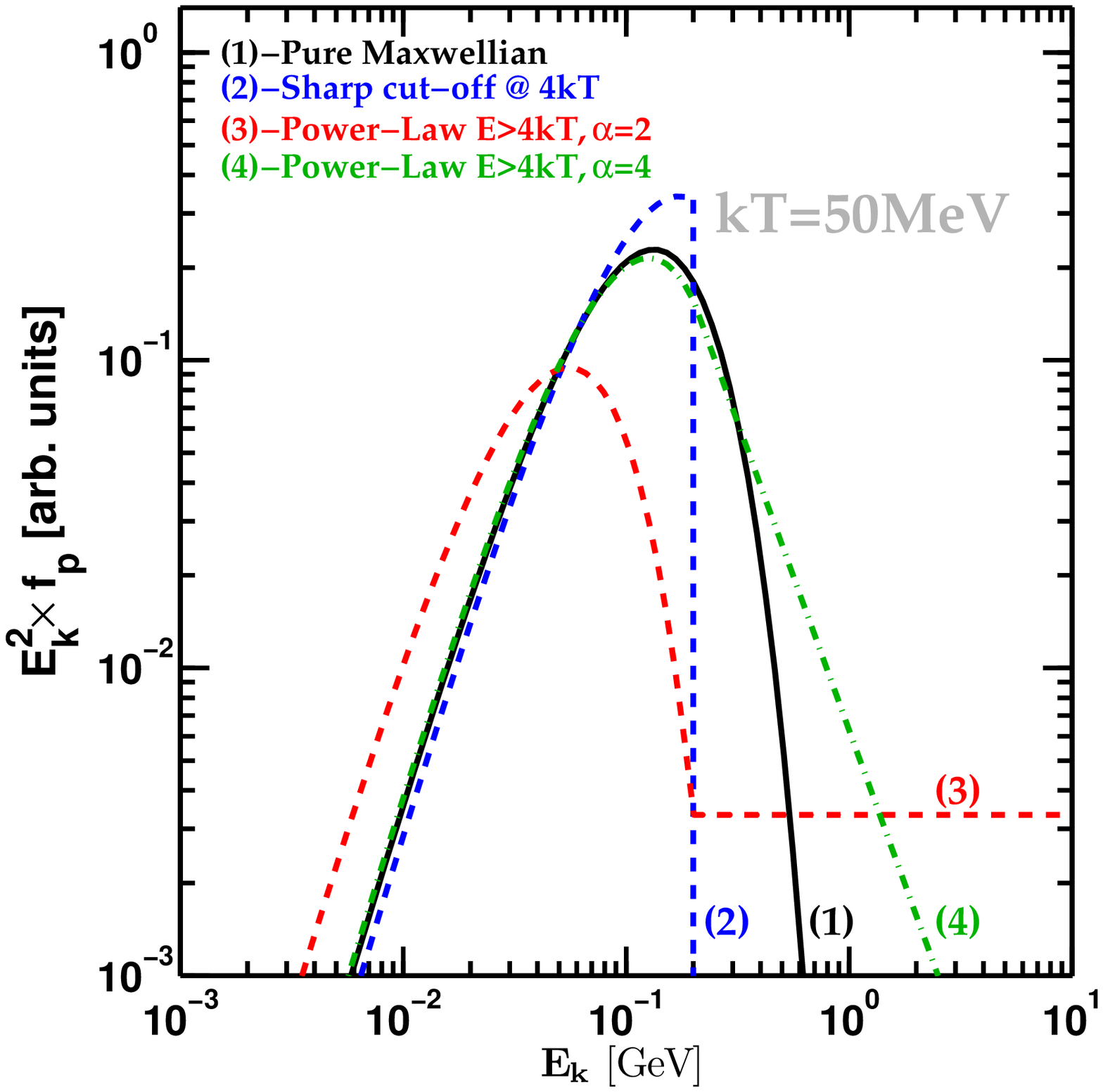}
\includegraphics[scale=0.4]{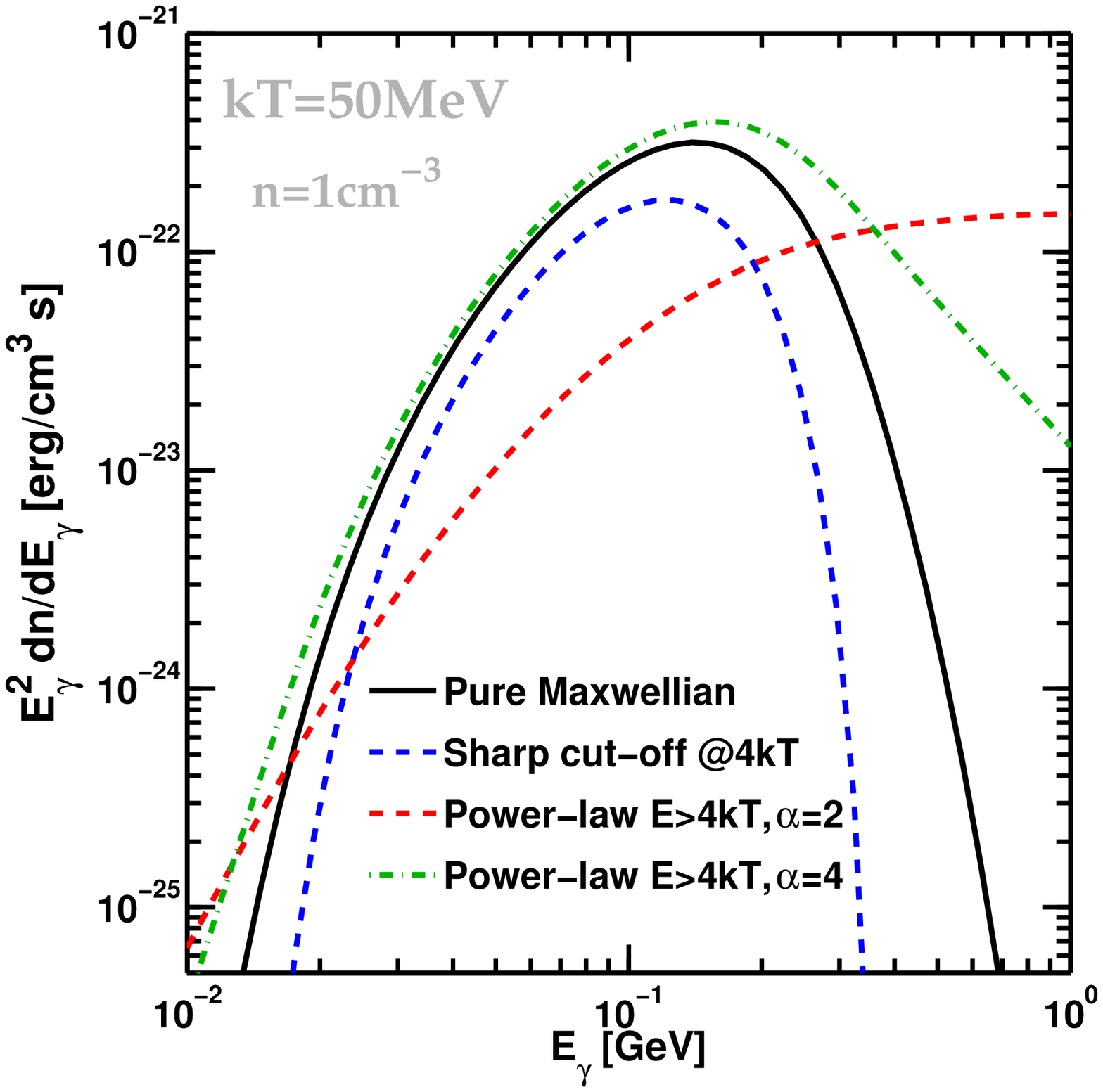}
\caption{Modified Maxwellian distribution functions and their $\pi^0$-decay $\gamma$-ray emissivities for two proton temperatures, $k T_{\rm p}=30$~MeV (top) and 50~MeV (bottom). {\it Left panels}: four proton distribution functions (Maxwellian, Maxwellian with a sharp cutoff at $4 k T_{\rm p}$, Maxwellian with power-law tails starting at $E=4k T_{\rm p}$ with indices $\alpha=2$ and 4). {\it Right panels}: the corresponding $\gamma$-ray luminosities. The plasma number density is set to $n_p=1~{\rm cm^{-3}}$. The distributions have been modified to conserve the total number of particles and the total energy of the original Maxwellian distribution as described in Eq.~(\ref{eq:distEcons}). \label{fig:ppSpecModif}}
\end{figure*}

Using the same formalism described above, we compute now the $\pi^0$-decay $\gamma$-ray emissivities for plasma with a modified Maxwellian distribution at high energies. Namely, we consider two type of deviations from the Maxwellian distribution at energies significantly exceeding the ion temperature: (1) Maxwellian distribution with a sharp cutoff at the kinetic energy $E_k^{\rm cut}=4\,k T_{\rm i}$, and (2) Maxwellian distribution with a power-law tail that starts at $E_k=4\,k T_{\rm i}$ and continues with a power-law index (a) $\alpha=2$ (b) $\alpha=4$. Together with the pure thermal Maxwellian distribution, these modified proton distribution functions (in arbitrary units) are shown in the left panels of Fig.~\ref{fig:ppSpecModif} for two proton temperatures $k T_{\rm p}=30$ and 50~MeV. The right panels show the corresponding $\gamma$-ray emissivities. By definition, the proton number density in these calculations should be set $n=1\,{\rm cm^{-3}}$, i.e. the proton distributions are normalized. Also the temperatures are modified in order to conserve the total energy of the original Maxwellian distribution (see Eq.~(\ref{eq:distEcons})).

As expected, the sharp cutoff in the region of the Maxwellian tail results in a strong suppression of $\gamma$-ray emission at all energies of $\gamma$-rays. The suprathermal power-law distributions above the Maxwellian distribution result in a strong modification of $\gamma$-ray spectra. While at high energies the emissivity is dramatically enhanced, at energies below $k T_{\rm i}$, it can be enhanced or suppressed depending on the temperature. 

\section{Discussion}

The gamma-radiation of plasma with an ion temperature not much higher than 10~MeV is dominated by the prompt nuclear de-excitation lines. At such temperatures, the plasma undergoes a rapid chemical evolution. The first stage of evolution is characterized by destruction of nuclei: in particular, the CNO group, the major contributors to the prompt $\gamma$-ray line emission. This stage is also characterized by dramatic enhancement of the LiBeB group by the secondary fragments of the destruction of heavier elements. However, the production of new elements cannot prevent the massive destruction of all nuclei, including $^4$He. Eventually, this results in the formation of a neutron-proton plasma with traces of light elements, first of all, deuterium.

During the first stage of the chemical evolution, the $\gamma$-ray emission is dominated by the prompt nuclear lines, and, at very high temperatures, also by the $\pi^0$ decay $\gamma$-rays. Later, with the dramatic reduction of the content of nuclei, the emission of nuclear $\gamma$-ray lines is suppressed with a few relatively strong lines linked to reactions with the involvement of $\alpha$ particles: $\alpha+\alpha\to^7$Li(0.478 MeV) and $\alpha+\alpha\to^7$Be(0.431 MeV). The last stage of the neutron-proton plasma, the $\gamma$-ray spectrum is dominated by the $\pi^0$-meson production, with some contribution at low energies from the $n+p$ radiative capture and the $n+p$ bremsstrahlung.

The relative contributions of different channels strongly depend on the temperature. When changing the plasma temperature from $k T= 30$ to 100 MeV, the emissivity of the $\pi^0\to2\gamma$ channel increases by more than two orders of magnitude. Meanwhile, the nuclear $\gamma$-ray lines become less visible, in particular, due to the Doppler--broadening. For the initial solar composition of plasma, the emissivity of the strongest nuclear $\gamma$-ray lines (${\rm^{12}C(4.4)}$ and ${\rm^{16}O(6.1\&6.3)}$) and the $\pi^0$-decay $\gamma$-rays become comparable for plasma temperature around  50~MeV.

The limited sensitivity of $\gamma$-ray detectors in the MeV energy band, makes the detection of strongest $\gamma$-ray lines from hot astrophysical plasmas rather difficult. The situation could be changed considerably after the arrival of new generation detectors like eASTROGAM and AMEGO which can be served as unique tools for the search, detection and identification of these distinct carriers of information about hot two temperature plasmas and related high energy phenomena close to compact relativistic objects. The same instruments, designed for studies of $\gamma$-rays both in MeV and GeV bands, can be used for the exploration of the $\pi^0$-decay $\gamma$-ray continuum. The characteristic narrow spectral energy distribution (SED) with a maximum at 100~MeV substantially differs from the spectra of $\pi^0$-decay $\gamma$-rays from most of the nonthermal source populations. The $\gamma$-ray emission of these objects is produced by accelerated protons with broad (typically, power-law) distributions. Correspondingly, the SED of radiation of these objects is broad with maximum well above 100~MeV. Thus, the narrow SED around 100~MeV can be served as a distinct signature for the search of very hot astrophysical plasmas with temperature exceeding $10^{10}$K. Prime targets could be the accreting solar-mass black holes in our Galaxy \cite{kafexhiu2018a}. While searching for $\gamma$-radiation with spectra expected from the Maxwellian distribution of ions, one should keep in mind possible deviations of distributions of protons and nuclei from the ``nominal'' Maxwellian distribution. The {\it under-developed} Maxwellian tails in the proton distributions,  e.g. due to the short lifetime of plasma compared to the characteristic relaxation times, would significantly suppress the "$\pi^0$-decay" $\gamma$-ray emission down to the level below the sensitivity of detectors. On the other hand, the presence of the possible {\it supra-thermal tails} of the proton distributions would significantly increase the luminosity of $\gamma$-rays, and therefore the chance of their detection.

In summary, the results of this study represent the first detailed calculations of $\gamma$-ray signatures of hot two-temperature astrophysical plasmas based on the study of chemical evolution of elements. The  presented results have practical implications for the interpretation of future $\gamma$-ray observations of compact relativistic objects or explosive phenomena with the formation of very hot two-temperature plasmas.

\begin{acknowledgments}
The authors  are grateful to the members of the High Energy Astrophysics Theory Group for the very fruitful discussions, and appreciate the grant of the Russian Science Foundation 16-12-1044, the NSF grant AST-1306672, DoE grant DE-SC0016369, and NASA grant 80NSSC17K0757.
\end{acknowledgments}

\appendix
\section{$p+p\to\pi^0\to2\gamma$ production differential cross section for the plasma rest frame of reference \label{sec:Appendix}}

The invariant $\pi^0$-meson production differential cross section is (here we use the natural units, i.e. $\hbar=c=k=1$):
\begin{equation}\label{Aeq:1}
E_\pi\,\frac{d^3\sigma}{dp_\pi^3}=\frac{1}{P_\pi}\,\frac{d^3\sigma}{dE_\pi d\Omega},
\end{equation}
\noindent where, $E_\pi$ and $P_\pi$ are the energy and momentum of the $\pi^0$-meson, respectively; $\Omega$ is the emission solid angle. Low collision energy experimental data at the 
center-of-mass reference frame show that $\pi^0$-mesons are produced almost isotropically. Therefore, Eq.~(\ref{Aeq:1}) for the center-of-mass reference frame transforms into $E_\pi\,d\sigma/dp_\pi^3=(4\pi P_\pi)^{-1}\,d\sigma/dE_\pi$, where, the $d\sigma/dE_\pi$ is measured experimentally. If we denote with prime the quantities in the center-of-mass frame, the energy distribution of $\pi^0$-mesons in an arbitrary frame of reference is given by:
\begin{equation}\label{Aeq:2}
\frac{d\sigma}{dE_\pi} = \frac{P_\pi}{2}\,\int\limits_{\mu_{\rm min}}^1 \frac{1}{P_\pi '}\,\frac{d\sigma '}{dE_\pi '}\,d\mu,
\end{equation}
where, $\mu=\cos(\theta_\pi)$. Let the two protons have Lorentz factors $\gamma_1$ and $\gamma_2$, and speeds $\beta_1$ and $\beta_2$, respectively. The center of mass energy squared $s = 2m_p^2(1+\gamma_r)$, where, the $\gamma_r=\gamma_1\,\gamma_2\,\left(1-\beta_1\,\beta_2\,\cos\alpha\right)$ is the collision Lorentz factor and $\alpha$ is the angle between the two colliding protons. The center of mass Lorentz factor is $\gamma_c=m_p(\gamma_1+\gamma_2)/\sqrt{s}$. The relation between the pion center of mass frame energy (here noted with prime) and the laboratory frame is given by:
\begin{equation}\label{Aeq:3}
E_\pi '= \gamma_c\,(E_\pi - \beta_c\,P_\pi\,\mu).
\end{equation}
Therefore, the minimum $\mu$ allowed by the kinematics is when the pion energy in the center-of-mass frame reaches its maximum possible value $E_\pi^{\rm max'}=(s-4m_p^2 +m_\pi^2)/(2\sqrt{s})$. Using Eq.~(\ref{Aeq:3}) we derive:
\begin{equation}
\mu_{\rm min}=\frac{\gamma_c\,E_\pi - E_\pi^{\rm max'}}{\gamma_c\,\beta_c\,P_\pi}
\end{equation}

Adopting Eq.~(\ref{Aeq:2}) and the kinematic transformation of the $\pi^0\to2\gamma$ decay we can calculate the $\gamma$-ray production cross section
\begin{equation}
\frac{d\sigma}{dE_\gamma} =2\times\int\limits_{Y_\gamma}^{E_\pi^{\rm max}} \frac{d\sigma}{dE_\pi} \;\frac{dE_\pi}{P_\pi}.
\end{equation} 
Here, $E_\pi^{\rm max}=\gamma_c(E_\pi^{\rm max'}+ \beta_c\,P_\pi^{\rm max'})$ is the maximum pion energy in the laboratory frame, and $Y_\gamma = E_\gamma + m_\pi^2/(4\,E_\gamma)$, with $E_\gamma$ as the observed $\gamma$-ray energy.

Using the center-of-mass reference frame parametrization of the pion energy distribution \cite{kafexhiu2014} and the appropriate kinematic transformations, one obtains the $d\sigma/dE_\gamma$ for the laboratory frame. To simplify numerical calculations, we have parametrized it as follows:
\begin{equation}
\frac{d\sigma}{dE_\gamma} = A_{\rm max}(\gamma_r,\gamma_c) \times F(\gamma_r,\gamma_c,Y_ \gamma). 
\end{equation} 
 $A_{\rm max}(\gamma_r,\gamma_c)$ is the peak value of $d\sigma/dE_\gamma$, whereas, $F(\gamma_r,\gamma_c,Y(E_\gamma))$ is the only term that depends on the $\gamma$-ray energy.
It describes the spectral shape and varies between 0 and 1. Most of the $d\sigma/dE_\gamma$ parametrizations in the literature are obtained in the frame where one of the colliding protons is at rest. The center-of-mass Lorentz factor for this case is $\gamma_c=\gamma_{c0}=\sqrt{(\gamma_r+1)/2}$. Let us denote with $\gamma_\pi^{\rm max}=E_\pi^{\rm max}/m_\pi$, the maximum pion Lorentz factor in the laboratory frame. The maximum $\gamma$-ray energy is obtained by $E_\gamma^{\rm max/min}=m_\pi\,\gamma_\pi^{\rm max}(1\pm\beta_\pi^{\rm max})/2$. In addition, let us denote with
\begin{equation}
\Gamma_c = \frac{\gamma_c\,\beta_c}{\gamma_{c0}\,\beta_{c0}}~~\text{and}~~
X_\gamma = \frac{Y_\gamma-m_\pi}{Y_\gamma^{\rm max} -m_\pi}\in [0,1],
\end{equation}
where, $\gamma_c$ and $\beta_c$ are the center-of-mass frame Lorentz factor and speed in the laboratory frame and $\gamma_{c0}$ and $\beta_{c0}$ are the same quantities but for the specific frame where one of the colliding protons is at rest. The quantity $Y_\gamma^{\rm max}$ is calculated from $Y_\gamma^{\rm max} = E_\gamma^{\rm max} + m_\pi^2/(4E_\gamma^{\rm max})$.

Let us define the $\xi_p=E_k/m_p=(\gamma_r-1)$, where, $E_k$ is the protons collision kinetic energy and $m_p$ is the proton mass. The parametrization of $A_{\rm max}(\gamma_r,\gamma_c)$ and $F(\gamma_r,\gamma_c,Y_\gamma)$ for $E_k<1$~GeV are: 

\begin{equation}
\begin{split}
A_{\rm max}(\gamma_r,\gamma_c) &= 5.9\times\frac{\sigma_\pi(\gamma_r)}{E_\pi^{\rm max0}}\times\left(A_1(\xi_p)\,\Gamma_c^{A_2(\xi_p)}+A_3(\xi_p) \right)\\
F(\gamma_r,\gamma_c,E_\gamma) &= \left(1-X_\gamma\right)^{\kappa(\xi_p)}.
\end{split}
\end{equation}
Here $E_\pi^{\rm max0}=\gamma_{c0}(E_\pi^{\rm max'}+ \beta_{c0}\,P_\pi^{\rm max'})$ is the pion maximum energy in the case when one of the protons is at rest; $\sigma_\pi$ is the $p+p\to\pi^0$ inclusive production cross section. The $A_1$, $A_2$, $A_3$ and $\kappa$ functions are defined as 
\begin{equation}
\begin{split}
A_1(\xi_p)&=\xi_p^{3/2}\times\left(2.64\,\xi_p-0.52\right)^{-1}\\
A_2(\xi_p)&=1.116 - 0.056\,\xi_p^{-1}\\
A_3(\xi_p)&=1.01 - A_1(\xi_p)\\
\kappa(\theta)&=3.29 - \frac{1}{5}\,\xi_p^{-3/2}\\
\end{split}
\end{equation}

For $1\leq E_k \leq 5$~GeV, the parametrization transforms as follows:
\begin{equation}
\begin{split}
A_{\rm max}(\gamma_r,\gamma_c) &= \alpha_3\,\xi_p^{\alpha_2}\exp\left(\alpha_1\,\theta^2\right)\frac{\sigma_\pi}{m_p}\times \frac{B_1(\xi_p)\,\Gamma_c +B_2(\xi_p) }{\Gamma_c +B_3(\xi_p)}\\
F(\gamma_r,\gamma_c,E_\gamma) &= \frac{\left(1-X_\gamma\right)^{\delta(\xi_p)}}{\left(1+X_\gamma/G\right)^{\epsilon(\xi_p)}}.
\end{split}
\end{equation}
Here, $\sigma_\pi$ is the $p+p\to\pi^0$ inclusive production cross section and functions $B_1$, $B_2$, $B_3$, $\delta$, $\epsilon$ and $G$ are defined as follows:

\begin{equation}
\begin{split}
B_1(\xi_p)&= \xi_p\times \left(0.645\,\xi_p-0.376\right)^{-1},\\
B_2(\xi_p)&= \xi_p^{3/4}\times \left(0.695\,\xi_p - 0.493\right)^{-1},\\
B_3(\xi_p)&= \xi_p^{3/4}\times \left(0.414\,\xi_p - 0.277\right)^{-1},\\
\delta(\xi_p)&= \mu(\xi_p)+2.45,\\
\epsilon(\xi_p)&= \mu(\xi_p)+1.45,\\
G &=3\times\frac{m_\pi}{Y_\gamma^{\rm max}},\\
\mu(q)&= \frac{5}{4}\,q^{5/4}\,\exp\left(-\frac{5}{4}\,q\right).\\
\end{split}
\end{equation}
Here, $q=(E_k-1~{\rm GeV})/m_p$ and $\alpha_1=0.054$, $\alpha_2=-0.52$ and $\alpha_3=9.53$.

\bibliographystyle{apsrev}
\bibliography{citation}

\end{document}